\documentclass{article}

\usepackage[utf8]{inputenc} % allow utf-8 input
\usepackage[T1]{fontenc}    % use 8-bit T1 fonts
\usepackage{hyperref}       % hyperlinks
\usepackage{url}            % simple URL typesetting
\usepackage{booktabs}       % professional-quality tables
\usepackage{amsfonts}       % blackboard math symbols
\usepackage{nicefrac}       % compact symbols for 1/2, etc.
\usepackage{microtype}      
\usepackage{lipsum}
\usepackage{graphicx}
\usepackage{indentfirst}
\usepackage{amsmath}
\usepackage{amsthm}
\usepackage{amssymb}
\usepackage{lineno}
\usepackage{setspace}
\usepackage{lineno}
\usepackage{authblk}
\graphicspath{ {./images/} }
\usepackage{geometry}
\title{\textbf{State-clustering method of payoff computation in repeated multiplayer games}}

\author[1]{Fang Chen \footnote{These authors contributed equally.}}
\newcommand\CoAuthorMark{\footnotemark[\arabic{footnote}]}
\author[2]{Te Wu \protect\CoAuthorMark}
\author[1]{Guocheng Wang}
\author[1,3]{Long Wang \thanks{Corresponding author: Long Wang(longwang@pku.edu.cn)}}
\affil[1]{\footnotesize Center for Systems and Control, College of Engineering, Peking University, Beijing, China}
\affil[2]{Center for Complex Systems, Xidian University, Xi'an, China}
\affil[3]{Center for Multi-Agent Research, Institute for Artificial Intelligence, Peking University, Beijing, China}

\begin{document}
\maketitle
%\linenumbers 
\begin{abstract}
Direct reciprocity is a well-known mechanism that could explain how cooperation emerges and prevails in an evolving population. Numerous prior researches have studied the emergence of cooperation in multiplayer games. However, most of them use numerical or experimental methods, not theoretical analysis. This lack of theoretical works on the evolution of cooperation is due to the high complexity of calculating payoffs. In this paper, we propose a new method, namely, the state-clustering method to calculate the long-term payoffs in repeated games. Using this method, in an $n$-player repeated game, the computing complexity is reduced from $O(2^n)$ to $O(n^2)$, which makes it possible to compute a large-scale repeated game’s payoff. We explore the evolution of cooperation in both infinitely and finitely repeated public goods games as an example to show the effectiveness of our method. In both cases, we find that when the synergy factor is sufficiently large,
the increasing number of participants in a game is detrimental to the evolution of cooperation. Our work provides a theoretical approach to study the evolution of cooperation in repeated multiplayer games.
\end{abstract}

% keywords can be removed


\section{Introduction}
Cooperation is vital for members’ survival and prosperity in living systems \cite{nowak2006five,kennedy2005don,nowak2006evolutionary,su2019evolutionary,li2020evolution,zhou2021aspiration,nowak1998evolution,hauert2002volunteering,henrich2006costly}. But for an individual, cooperation means he needs to pay a cost and benefit his opponent. If there is no other mechanism, cooperation is expected to perish. The past decades have seen numerous studies about this cooperation conundrum. Many powerful mechanisms were revealed, such as direct reciprocity \cite{nowak1993strategy,trivers1971evolution,pinheiro2014evolution,press2012iterated}, indirect reciprocity \cite{nowak1998evolution,Nowak2005,Hilbe2018}, and spatial reciprocity \cite{Nowak1992,Ohtsuki2006}. The most well-known mechanism, direct reciprocity, is based on repeated games. It captures the fact that two individuals may interact for many rounds, which thus enables one to reciprocate the opponent in the future. Many famous strategies were also found, such as TFT (Tit-for-Tat) \cite{trivers1971evolution}, WSLS (Win Stay, Lose Shift) \cite{nowak1993strategy}, AoN (All-or-None) \cite{pinheiro2014evolution}, and ZD (Zero-Determinant) strategies \cite{press2012iterated}.

When studying direct reciprocity, the most important process is to compute the payoff. For the simplest case, two-player two-action games (e.g. the repeated prisoner’s dilemma), the computation is greatly simplified if one assumes that the player’s action only depends on the last round, which is also known as memory-1 strategies. In each round, each player has two choices of action, cooperating (C) and defecting (D). Thus, there are four possible outcomes in each round, i.e., CC, CD, DC, DD. Each player behaves on the condition of the outcome of the last round. Thus it forms a Markov chain with four states. Traditionally, we need to compute the stationary distribution of this Markov chain to calculate players’ payoffs. However, for a multiplayer game, although one assumes each player still has two choices of action, the corresponding Markov chain has $2^n$ states for an $n$-player game. The number of states explodes exponentially as the number of players increases. Due to this complexity of computation, most of the prior studies about the evolution of cooperation in repeated multiplayer games are based on simulations \cite{hauert1997effects,hilbe2015evolutionary} or experiments \cite{yang2013nonlinear}, not theoretical analysis. Hilbe et al also explained in \cite{hilbe2014cooperation}: “This lack of research may be due to the higher complexity: the mathematics of repeated n-player dilemmas seems to be more intricate, and numerical investigations are impeded because the time to compute payoffs increases exponentially in the number of players.”

In this paper, we propose a state-clustering method to calculate the long-term payoffs in repeated multiplayer games. Using this method, the time of computing payoffs decreases drastically. Traditionally, for an $n$-player game, the computing time explodes exponentially as $n$ grows. However, our method reduces it to a square rate. We use this method to explore the evolution of cooperation in repeated multiplayer public goods games as an example. In both finitely and infinitely repeated games, large groups hinder the evolution of cooperation. Using this example, we would like to open the door to the evolution of cooperation in multiplayer games.

\section{Model}
\label{sec:model}

Here we consider a population with $N$ players. In each time step, $n$ individuals are selected to play a multiplayer game. Each individual has two choices: cooperating (C) and defecting (D). We assume that the game is symmetrical, which means one's payoff only depends on how many players cooperate and how many players defect. Table.~\ref{table1} presents the payoff structure. When there are $k$ cooperators among the $n-1$ co-players, a cooperator can obtain $a_k$ and a defector can obtain $b_k$. For a special case of multiplayer games, linear Public Goods Game (PGG), each player can decide whether to contribute an endowment $c$ to the common pool, and the total endowment will be multiplied by a synergy factor $r$ $(1<r<n)$ and then evenly divided among $n$ players. We call the player who contributes an endowment $c$ cooperator and who does not contribute defector. Thus, we can get $a_k=(k+1)rc/n-c$ and $b_k=krc/n$. For arbitrary $k$, a cooperator's payoff is always lower than a defector's, so every rational player will choose defecting, which forms a Nash equilibrium.  

After one round of interaction, another round occurs with probability $\delta$ $(0<\delta \le 1)$. A player's strategy describes how the player acts according to the outcomes of previous rounds. Here we consider the well-known memory-1 strategies. Players' decision of choosing cooperating or defecting only depends on the previous round. Due to the symmetry, each player only considers the number of cooperators and his own action in the last round. For example, when player $A$ cooperates, and there are $k$ cooperators and $n-1-k$ defectors among the co-players in the last round, player $A$ will cooperate in this round with probability $p_{Ck}$ and defect with probability $1-p_{Ck}$. In particular, a player will cooperate with probability $p_0$ in the initial round. Thus, a player's strategy can be written as a vector $\mathbf{p}=[p_0, p_{C0}, p_{C1}, ..., p_{Cn-1}, p_{D0}, p_{D1}, ..., p_{Dn-1}]$. Furthermore, the execution error called ``trembling hands'' is also considered in our model. With probability $\epsilon$, a player originally intending to cooperate (defect) chooses to defect (cooperate) by mistake. So, a strategy can be transformed to its effective strategy by $\mathbf{\tilde{p}}=(1-\epsilon) \mathbf{p}+\epsilon (1-\mathbf{p})$. All the following strategies in this paper have already been transformed unless otherwise specified.

Each player's payoff is the average payoff over all rounds. If $\delta=1$, we say this game is infinitely repeated, then a player's long-term payoff is
\begin{equation}
    \pi=\lim_{T\to \infty}\frac{1}{T} \sum_{t=1}^T \pi(t),
\end{equation}
where $\pi (t)$ is the player's payoff in round $t$. If $0<\delta<1$, this game is finitely repeated. The game will proceed $1/(1-\delta)$ rounds on average. Thus, a player's long-term payoff is
\begin{equation}
    \pi=(1-\delta)\sum_{t=1}^\infty \delta^t \pi(t).
\end{equation}
\begin{table}[ht]
    \centering
    \caption{\textbf{Payoffs of symmetrical multiplayer games.} For a symmetrical multiplayer game, the payoff of each player depends on his own action and the number of cooperators among his co-players. If $k$ co-players cooperate, a cooperator will get $a_k$ and a defector will get $b_k$.}
        \begin{tabular}{c|cccccc}
    \hline
         \textbf{number of cooperators among co-players} & 0 & 1 & $\cdots$ & $k$ & $\cdots$ & $n-1$  \\ \hline
         cooperator's payoff & $a_0$ & $a_1$ & $\cdots$ & $a_k$ & $\cdots$ & $a_{n-1}$  \\ \hline
         defector's payoff & $b_0$ & $b_1$ & $\cdots$ & $b_k$ & $\cdots$ & $b_{n-1}$ \\
         \hline
    \end{tabular}
    \label{table1}
\end{table}

\section{Results}
\label{sec:results}

\paragraph{The state-clustering method of calculating payoffs.} Consider a multiplayer game with $n$ players. Each player has two actions: cooperation (C) and defection (D). Thus, there are $2^n$ action profiles. Each action profile is a Markov chain's state. The transition probability from one state to another is depicted by every player's strategy. Traditionally, to get every player's payoff, we need to calculate the stationary distribution of this Markov chain, whose probability transition matrix $\mathbf{M}$ is $2^n$ dimensional. As the number of players increases, the time of computation explodes exponentially, which makes computation intractable.

\begin{figure}[h!]
    \centering
    \includegraphics[scale=0.5]{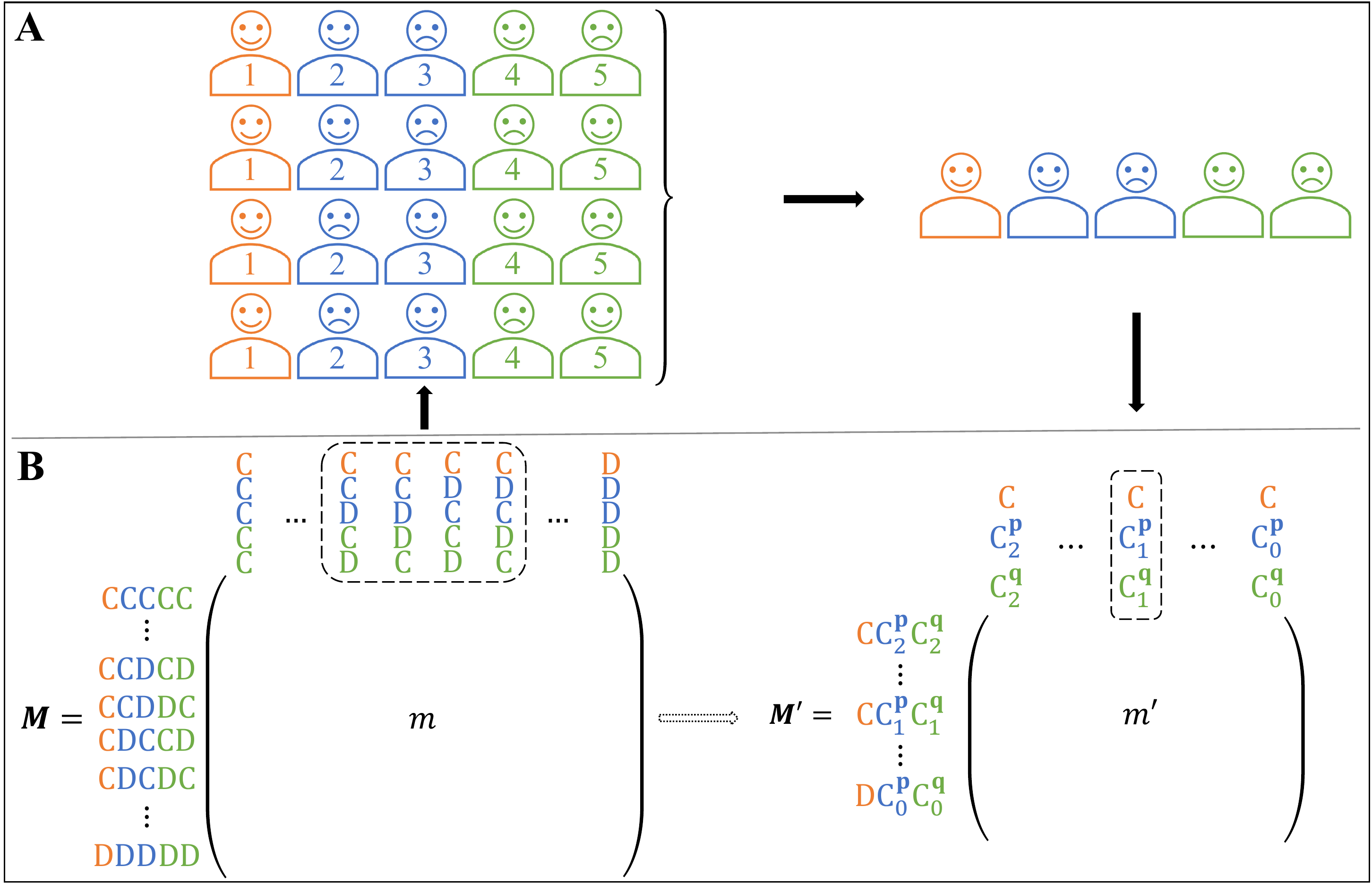}
    \caption{\textbf{An illustration of the reduced transition matrix.} Take the repeated game with five players as an example. Suppose two players adopt $\mathbf{p}$ ``blue'' players) and two players adopt $\mathbf{q}$ (``green'' players) in addition to the focal player (``orange'' player). Traditionally, we record all players' actions in a round as a state. Each player can choose cooperation or defection, so there are $2^n$ states in total. However, some states are undifferentiated. For example, there are four states which both have one cooperator among $\mathbf{p}$ players and one cooperator among $\mathbf{q}$ players. The transition probabilities from these four states to any states are equal. Thus, we can combine these four outcomes into one state (A). The dimension of the transition matrix slumps after this combination (B).}
    \label{model}
\end{figure}

Here, we discover a new method which could efficiently compute the players' payoffs in repeated multiplayer games, making it possible to theoretically study the evolution of cooperation. In evolution, if mutations are rare, there are at most two strategies in the population \cite{wu2018coevolutionary,wu2019phenotype,nowak2004emergence}. Thus, here we only consider the payoff when there are two strategies. The case of multiple strategies is investigated in Supplementary Information. If we only consider two strategies $\mathbf{p}$ and $\mathbf{q}$, due to the symmetry, some states actually have the same property. In each round, each player can choose cooperation or defection, and players can also be categorized by strategies into two types: $\mathbf{p}$ players and $\mathbf{q}$ players. Thus, there are four types of players in each round: $\mathbf{p}$ player who cooperates, $\mathbf{p}$ player who defects, $\mathbf{q}$ player who cooperates, and $\mathbf{q}$ player who defects. As long as two states have the same number of these four types, the two states are  undifferentiated. It means the Markov chain can be simplified. For convenience, we set the player whose payoff we want to compute as the focal player. For example, although the states
\begin{equation}
  \underbrace{C}_{\text {focal player }} \underbrace{C D D D D}_{\text {\textbf{p} players }} \underbrace{C C C D}_{\text {\textbf{q} players }}
  \label{equ:state1}
\end{equation}
and
\begin{equation}
  \underbrace{C}_{\text {focal player}} \underbrace{D D C D D}_{\text {\textbf{p} players }} \underbrace{C D C C}_{\text {\textbf{q} players}}
  \label{equ:state2}
\end{equation}  
are different, the payoff of the focal player is the same in this round. In other words, the focal player does not care who cooperates. He only cares how many $\mathbf{p}$ players cooperate and how many $\mathbf{q}$ players cooperate. Thus, to derive the payoff of a focal player, we can merge these states which have the same number of cooperators among $\mathbf{p}$ players and $\mathbf{q}$ players into one state (Fig.~\ref{model}). In a game with $n$ players, we suppose there are $k$ players who adopt strategy $\mathbf{p}$. Then, a reduced Markov chain's state consists of three elements: the action of the focal player, the number of cooperators among $\mathbf{p}$ players, and the number of cooperators among $\mathbf{q}$ players. We use $AC_x^{\mathbf{p}}C_y^{\mathbf{q}}$ to denote a Markov state, where $A\in\{C,D\}$. It means the focal player adopts $A$ and there are $x$ cooperators among $\mathbf{p}$ players and $y$ cooperators among $\mathbf{q}$ players.

 By doing this reduction, the number of the Markov chain's states is reduced from $2^n$ to $2(k+1)(n-k)$, varying with the number of cooperators present in the group. To give the transition probability between two states, we first define a map $f:\{C,D\}\to \{0,1\}$, which satisfies $f(C)=1$ and $f(D)=0$. If we want to calculate a $\mathbf{p}$ player's payoff, we just set that the focal player adopts strategy $\mathbf{p}$. Then, the transition probability from state $AC_x^{\mathbf{p}}C_y^{\mathbf{q}}$ to $A'C_{x'}^{\mathbf{p}}C_{y'}^{\mathbf{q}}$ can be calculated (see Methods). That is
\begin{equation}
  \begin{array}{ll}
    \left|1-f(A')-p_{Ax+y}\right| 
    &\left[\sum_{j=0}^{x^{\prime}} \tbinom{x}{j}p_{C(x+y-1+f(A))}^{j}\left(1-p_{C(x+y-1+f(A))}\right)^{x-j}\right. \\
    &\left.\tbinom{k-x}{x'-j} p_{D(x+y+f(A))}^{x^{\prime}-j}\left(1-p_{D(x+y+f(A))}\right)^{(k-x)-\left(x^{\prime}-j\right)}\right] \\
    &{\left[\sum_{j=0}^{y^{\prime}}\tbinom{y}{j} q_{C(x+y-1+f(A))}^{j}\left(1-q_{C(x+y-1+f(A))}\right)^{y-j}\right.} \\
    &\left.\tbinom{n-k-1-y}{y'-j} q_{D(x+y+f(A))}^{y^{\prime}-j}\left(1-q_{D(x+y+f(A))}\right)^{(n-1-k-y)-\left(y^{\prime}-j\right)}\right]
    \end{array}.
    \label{equ:transitionProbability1}
\end{equation}
When we want to calculate $\mathbf{q}$ player's payoff, we just set that the focal player adopts strategy $\mathbf{q}$, and the transition probability is similar. Compared to computing a stationary distribution of a Markov chain with $2^n$ states, a Markov chain with $2(k+1)(n-k)$ states is much more tractable (Fig.~\ref{model}). 
Collecting all transition probabilities, we get the transition matrix $\mathbf{M'}$. For non-zero $\epsilon$ and $\delta$, there is a unique average distribution $\mathbf{v}$ over all states. We denote the marginal distribution of state $AC_x^{\mathbf{p}}C_y^{\mathbf{q}}$ as $v_{AC_x^{\mathbf{p}}C_y^{\mathbf{q}}}$. Then, the payoff of the focal player (see Methods) is
\begin{equation}
  \pi = \sum_{x,y} v_{C C_{x}^{\mathbf{p}} C_y^{\mathbf{q}}} a_{x+y} + \sum_{x,y} v_{D C_{x}^{\mathbf{p}} C_y^{\mathbf{q}}} b_{x+y},
  \label{equ:payoff}
\end{equation}
where $\mathbf{v}=(1-\delta)\mathbf{v}(0)[\mathbf{I}-\delta\mathbf{M'}]^{-1}$. Particularly, if $\delta=1$, $\mathbf{v}$ is the stationary distribution of $\mathbf{M'}$, i.e., $\mathbf{vM'}=\mathbf{v}$. In Supplementary Information, we have proved our method is equivalent to the traditional method. 

\paragraph{Evolution of cooperation in repeated public goods games.} Traditionally, the evolution of repeated multiplayer games is often studied by simulations or experiments. Using our method, we can explore the evolution of cooperation in this case theoretically. We consider a mutation-selection process \cite{nowak2006evolutionary} in a population consisting of $N$ players. In each time step, two players $X$ and $Y$ are randomly chosen. Player $X$ has a chance to update his strategy. With probability $\mu>0$, mutations occurs and $X$ adopts an arbitrary memory-1 strategy equiprobably. With supplementary probability $1-\mu$ no mutation take place. Player $X$ imitates $Y$'s strategy with probability 
\begin{equation}
    \left(1+\exp \left[s\left(P_{X}-P_{Y}\right)\right]\right)^{-1},
\end{equation}
where $s$ is the selection intensity. Larger $s$ means payoffs are more important for the success of evolution.

When mutations are rare, a mutant will take over the whole population or become extinct before another mutation occurs. Thus, there are at most two strategies among the population. The probability that a mutant invades and takes over the whole population is called fixation probability \cite{nowak2004emergence}, denoted by $\rho$. We explore the evolution of cooperation using the stochastic evolutionary dynamics of Imhof and Nowak \cite{imhof2010stochastic}. Initially, all individuals are set to adopt a strategy randomly picked up from the memory-1 strategy space. After that, a mutant which is uniformly drawn from the space of memory-1 strategies is introduced to the population. The mutant will either take over the whole population with probability $\rho$ or become extinct with probability $1-\rho$. Then, the population becomes homogeneous again. At this time, another mutant is introduced to attempt to invade the whole population and the algorithm enters the next cycle.

\begin{figure}[t]
  \centering
  \includegraphics[scale=0.7]{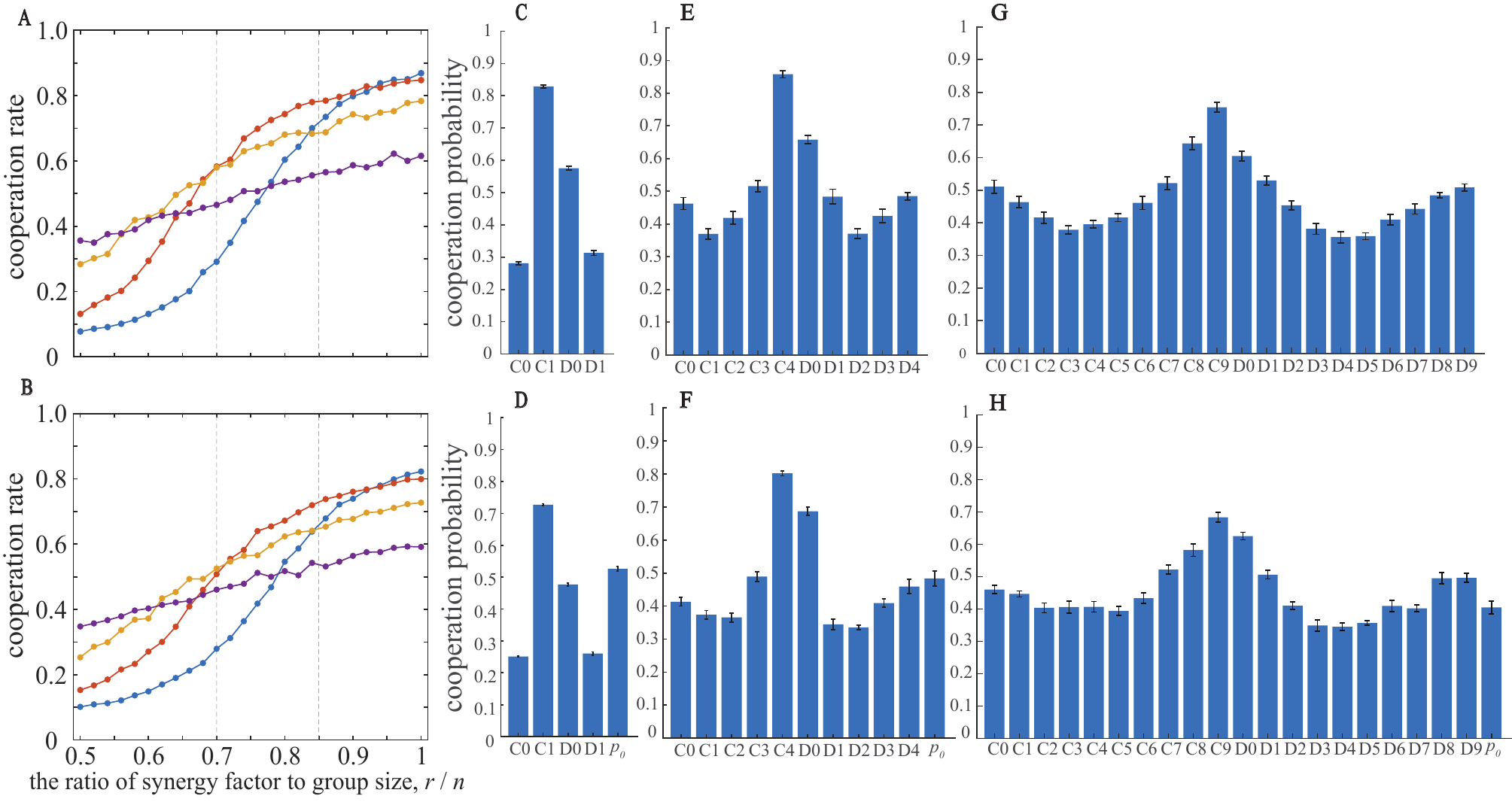}
  \caption{\textbf{Evolution of cooperation in repeated Public Goods Game.} We consider two scenarios: infinitely repeated Public Goods Game (upper) and finitely repeated Public Goods Game with $\delta=0.85$ (lower). To study the impact of group size on the evolution of cooperation, we used the stochastic evolutionary dynamics of Imhof and Nowak \cite{imhof2010stochastic}. Each mutant strategy is uniformly chosen from the memory-1 strategy space. We introduce $10^6$ mutant strategies to each simulation and performed 10 independent runs. (A, B) When the synergy factor is small, large group can promote the evolution of cooperation. However, when the synergy factor is large, large groups do not support the evolution of cooperation. (C-H) The average strategies during simulations for $n=2$ (C, D), $n=5$ (E, F) and $n=10$ (G, H). Color bars show the average cooperation probabilities and error bars portray the standard error. In all cases, players would cooperate with a high probability if most players cooperate or defect in the previous round. Players defect with a high probability if a medium number of players cooperate in the previous round. Parameters: $c=1$, $r/n=0.85$, $N=100$, $\epsilon=0.01$ and $s=1$.}
  \label{fig:cooperation}
\end{figure}

We first compute the cooperation rate (Fig.~\ref{fig:cooperation}A, B). Increase in $r/n$, the ratio
of synergy factor to group size, boosts cooperation for all group sizes but to different degrees. For large group size, cooperation rate rises but in a gentle rate. Raising $r/n$ can lead to substantial cooperation rate for small group size. This result implies that more than linearly increasing synergy factors is required to achieve the same cooperation rate as group size grows, further confirming that large group hinders the evolution of cooperation. 
Here we provide some intuitions: If the synergy factor is large, a small number of partners can bring considerable benefits to all players. But if the synergy factor is small, more participants are necessary in order to bring high benefits to all players. Thus, large $r$ gives players incentives to defect but small $r$ gives players incentives to cooperate.

\begin{figure}[ht]
  \centering
  \includegraphics[width=4.5in]{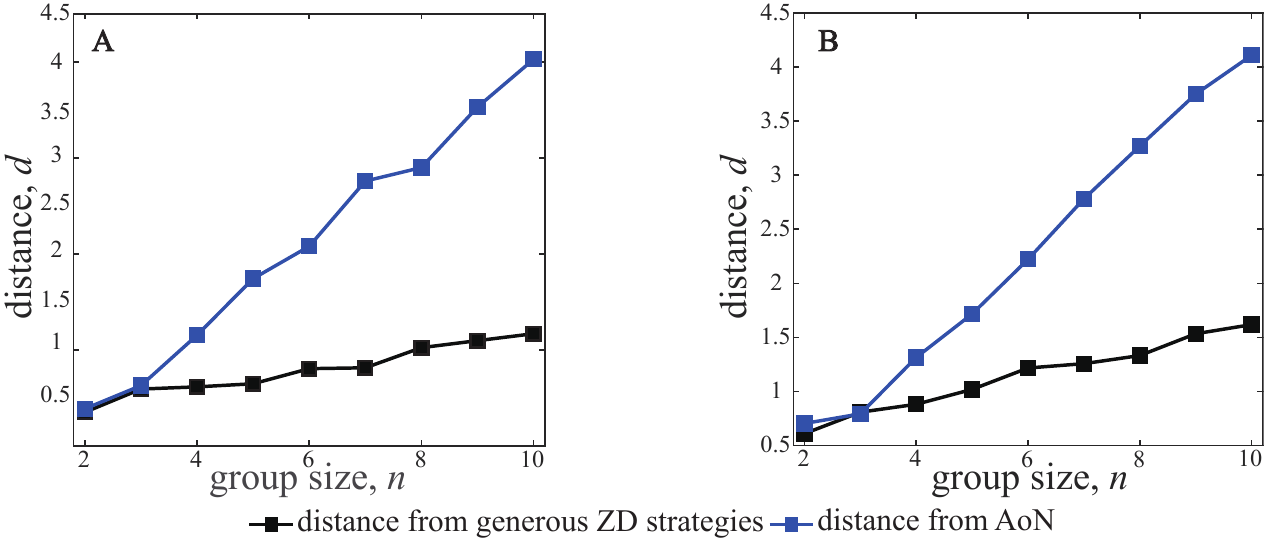}
  \caption{\textbf{Distances between the average strategies and some important memory-1 strategies.} Two scenarios are considered: infinitely repeated public goods game (A) and finitely repeated public goods game (B). We use the Euclidean distance to measure the similarity of the two strategies. In both cases, the distances between the average strategies and generous ZD strategies are closer than between AoN, especially in large groups. The parameters in (A) and (B) are the same as Fig.\ref{fig:cooperation}.}
  \label{fig:distance}
\end{figure}

We further compute the average resident strategies with 2, 5 and 10 players respectively (Fig.~\ref{fig:cooperation}C-H). In all cases, players tend to be highly cooperative when their co-players have carried out the same action in the previous round, and tend to be defective when their co-players choose different actions. This property is similar to All-or-None strategy to some extent. This result holds both in finitely or infinitely repeated games. However, some prior studies also show that in repeated games, generous ZD is the most robust strategy \cite{stewart2013extortion}, rather than All-or-None. Then we show the distance between the average resident strategies and the two well-known strategies. We use the squared Euclidean distance of two strategies to measure the similarity between them. Fig.~\ref{fig:distance} shows the average resident strategies' distances from generous ZD strategy and from All-or-None strategy. In both finitely and infinitely repeated games, the average resident strategy shows a higher level of similarity to generous ZD, consistent with the result reported in \cite{stewart2013extortion}. These results are also robust to discounted factors, synergy factors and distributions over which mutants are drawn from (Supplementary Information).

\section{Discussion}
Direct reciprocity is a well-known mechanism for the evolution of cooperation. Most prior studies concentrated on two-player games, such as the prisoners’ dilemma game, the snowdrift game. For multiplayer games, prior studies are mainly based on simulations or experiments. Theoretical analysis is little. This is due to the complexity of calculating payoffs. As the number of players increases, the complexity of calculating payoffs increases exponentially. 
 
Here, under the general assumption of symmetrical games \cite{hilbe2017memory,pan2015zero,hilbe2014cooperation}, we develop a new efficient method, which could reduce the computation complexity remarkably. When mutations are sufficiently rare, there are at most two strategies among the population, denoted by $\mathbf{p}$ and $\mathbf{q}$. Using symmetry, in each state, we only need to record the number of cooperators among $\mathbf{p}$ players and $\mathbf{q}$ players respectively. By doing this, we reduce the number of Markov chain’s states from $O(2^n)$ to $O(n^2)$, where $n$ is the number of players in a game. Furthermore, if there are $m$ strategies, the number of states is $O(n^m)$, which is still lower than $2^n$.  We performed simulations to compare the time of computing a 10-player repeated game in Fig. 2. Using the traditional methods, it takes at least two weeks, but our method only needs 7 hours. Thus, our method makes it possible to compute the payoffs in multiplayer games with large groups.
 
Using this method, we investigate the evolution of cooperation in repeated public goods games as an example. We find that the increasing number of participants in a public goods game hinders the evolution of cooperation. This result is also illustrated in \cite{boyd1988evolution}. In addition, previous findings show that the most abundant strategy in repeated public good games is All-or-None. However, in our findings, the most leading strategies are more similar to the generous ZD strategy. The difference is worthy of further study to elucidate.

This method gives an efficient way to compute the payoff and cooperation rate in multiplayer games. We hope this approach of payoff calculation could pave the way to exploring the evolutionary dynamics in repeated multiplayer games.

\section{Methods}\label{sec:method}
In the following, we provide a more technical summary of our method to calculate the long-term payoffs under the assumption of low mutations. The derivation without the assumption of low mutations is in Supplementary Information. This section consists of three parts: i) derivations of the reduced transition matrix, ii) how to use reduced transition matrix to calculate the long-term payoffs in both infinitely and finitely repeated multiplayer games, and iii) how to use reduced transition matrix to compute the cooperation rates.

\paragraph{Derivations of reduced transition matrix.} The transition matrix plays a critical role in calculating long-term payoffs of both discounted and undiscounted repeated multiplayer games. The transition matrix $\mathbf{M}$ collects all transition probabilities between any two states. Traditionally, each state is written as $\{(A_1,...,A_n)|A_i \in \{C,D\}\}$, where $A_i$ is the action of player $i$. Thus, the dimension of transition matrix $\mathbf{M}$ is $2^n$, which increases exponentially in the group size $n$. Computing payoff needs to calculate the left eigenvector of this matrix, which is intractable when the group size $n$ is large.

Under the limit of rare mutations, there are at most two different strategies $\mathbf{p}$ and $\mathbf{q}$ in the population, where $\mathbf{p}=[p_0, p_{C0}, ..., p_{Cn-1}, p_{D0}, ..., p_{Dn-1}]$ and $\mathbf{q}=[q_0, q_{C0}, ..., q_{Cn-1},q_{D0}, ..., q_{Dn-1}]$. Here, $p_{Ck}$ ($p_{Dk}$) denotes the probability that the focal player cooperates in this round when he cooperates (defects) and $k$ of his co-players cooperate in the last round. We emphasize that due to the symmetry, some states can be combined into one state. For a focal player, only the number of cooperators among his co-players can affect his payoff. Who cooperate and who defect does not matter. Thus, for a focal player, we use $AC_x^{\mathbf{p}}C_y^{\mathbf{q}}$ to denote a state. Here, $A\in\{C,D\}$ is the focal player's action, and $C_x^{\mathbf{p}}$ ($C_y^{\mathbf{q}}$) means there are $x$ ($y$) of his co-players who adopt $\mathbf{p}$ ($\mathbf{q}$) cooperating. 

Here, the states are constructed in the view of a focal player. The transition probability between two states is determined by the focal player's strategy and how many $\mathbf{p}$ and $\mathbf{q}$ co-players cooperate. Collecting all these transition probabilities, we obtain a reduced Markov chain. To compute the focal player's payoff, we need to calculate this Markov chain's stationary distribution. However, it does not means that we need to solve $n$ Markov chains if we want to get each player's payoff. Due to the symmetry, we only need to solve two Markov chains: the focal player adopts strategy $\mathbf{p}$ and $\mathbf{q}$.

Suppose $k$ of the focal player's co-players adopt strategy $\mathbf{p}$ and $n-1-k$ co-players adopt $\mathbf{q}$. Then, we will give the transition probability from state $AC_x^{\mathbf{p}}C_y^{\mathbf{q}}$ to $A'C_{x'}^{\mathbf{p}}C_{y'}^{\mathbf{q}}$. For convenience, we first introduce a function $f$ with $f(C)=1$ and $f(D)=0$. At state $AC_x^{\mathbf{p}}C_y^{\mathbf{q}}$, if the focal player adopts strategy $\mathbf{p}$, he will cooperate in the next round with probability $p_{A(x+y)}$ and defect with probability $1-p_{A(x+y)}$. The probability that the focal player's action changes from $A$ to $A'$ is 
\begin{equation}
    \label{equ: transition_1}
    |1-f(A')-p_{A(x+y)}|.
\end{equation}
Suppose there are $x$ cooperators among $\mathbf{p}$ co-players. The probability that the number of cooperators changes to $x'$ is 
\begin{equation}
    \label{equ: transition_2}
    \begin{split}
         \sum_{j=0}^{x^{\prime}} & \tbinom{x}{j}p_{C(x+y-1+f(A))}^{j}\left(1-p_{C(x+y-1+f(A))}\right)^{x-j} \\
    & \tbinom{k-x}{x'-j} p_{D(x+y+f(A))}^{x^{\prime}-j}\left(1-p_{D(x+y+f(A))}\right)^{(k-x)-\left(x^{\prime}-j\right)}
    \end{split}
\end{equation}
Suppose there are $y$ cooperators among $\mathbf{q}$ co-players. The probability that the number of cooperators changes to $y'$ is
\begin{equation}
    \label{equ: transition_3}
    \begin{split}
        \sum_{j=0}^{y^{\prime}}\tbinom{y}{j} & q_{C(x+y-1+f(A))}^{j}\left(1-q_{C(x+y-1+f(A))}\right)^{y-j} \\
    & \tbinom{n-k-1-y}{y'-j} q_{D(x+y+f(A))}^{y^{\prime}-j}\left(1-q_{D(x+y+f(A))}\right)^{(n-1-k-y)-\left(y^{\prime}-j\right)}
    \end{split}
\end{equation}
Thus, if the focal player adopts strategy $\mathbf{p}$, the transition probability from state $AC_x^{\mathbf{p}}C_y^{\mathbf{q}}$ to $A'C_{x'}^{\mathbf{p}}C_{y'}^{\mathbf{q}}$ is the product of Eq.~\ref{equ: transition_1}, Eq.~\ref{equ: transition_2} and Eq.~\ref{equ: transition_3}. Then, we obtain Eq.~\ref{equ:transitionProbability1}. Collecting all these transition probabilities, we get the reduced transition matrix $\mathbf{M'}$.

\paragraph{Long-term payoffs in infinitely and finitely repeated multiplayer games.} We have reduced the dimension of the transition matrix by combining certain states into one state. Now, we will show how to use the transition matrix to calculate payoffs.

Let's begin with infinitely repeated multiplayer games. With the assumption of a trembling hand, all states are ergodic. As a result, the reduced transition matrix $\mathbf{M}'$ has a unique left eigenvector $\mathbf{v}$ corresponding to eigenvalue 1. Each element of $\mathbf{v}$, $v_{A C_{x}^{\mathbf{p}} C_y^{\mathbf{q}}}$, is the probability the focal player finds himself in the state $A C_{x}^{\mathbf{p}} C_y^{\mathbf{q}}$. Therefore, the payoff of the focal player is 
\begin{equation}
  \pi = \sum_{x,y} v_{C C_{x}^{\mathbf{p}} C_y^{\mathbf{q}}} a_{x+y} + \sum_{x,y} v_{D C_{x}^{\mathbf{p}} C_y^{\mathbf{q}}} b_{x+y}.
  \label{equ:payoff}
\end{equation}

Then, we calculate long-term payoffs in finitely repeated multiplayer games. Let $\mathbf{v}$ denote the mean distribution of reduced states. Each element of $\mathbf{v}$, $v_{A C_{x}^{\mathbf{p}} C_y^{\mathbf{q}}}$ represents the probability that the focal player lies in the state $A C_{x}^{\mathbf{p}} C_y^{\mathbf{q}}$. The mean distribution $\mathbf{v}$ is given by
\begin{equation*}
  \mathbf{v} =\sum_{t=0}^{\infty}(1-\delta) \delta^{t} \mathbf{v}(t)
\end{equation*}
where $\mathbf{v}(t)$ represents the distribution of states at time $t$. We can calculate the explicit form of $\mathbf{v}$ according to
\begin{equation}
  \begin{aligned}
    \mathbf{v} &=\sum_{t=0}^{\infty}(1-\delta) \delta^{t} \mathbf{v}(t) \\
    &=(1-\delta) \mathbf{v}(0)+\sum_{t=1}^{\infty}(1-\delta) \delta^{t} \mathbf{v}(t) \\
    &=(1-\delta) \mathbf{v}(0)+\sum_{t=0}^{\infty}(1-\delta) \delta^{t+1} \mathbf{v}(t+1) \\
    &=(1-\delta) \mathbf{v}(0)+\delta \sum_{t=0}^{\infty}(1-\delta) \delta^{t} \mathbf{v}(t) \mathbf{M}' \\
    &=(1-\delta) \mathbf{v}(0)+\delta \mathbf{v} \mathbf{M}',
  \end{aligned}
  \label{equ:vbellman}
\end{equation}
which is known as the Bellman equation. Here, $\mathbf{v}(0)$ is the initial distribution of the reduced states. If  the focal player applies the strategy $\mathbf{p}$, the element of $\mathbf{v}$ satisfy $v_{CC_x^{\mathbf{p}}C_y^{\mathbf{q}}}(0)=p_0\tbinom{k}{x}p_0^x (1-p_0)^{k-x}\tbinom{n-k-1}{y}q_0^y (1-q_0)^{n-k-1-y}$ and $v_{DC_x^{\mathbf{p}}C_y^{\mathbf{q}}}(0)=(1-p_0)\tbinom{k}{x}p_0^x (1-p_0)^{k-x}\tbinom{n-k-1}{y}q_0^y (1-q_0)^{n-k-1-y}$. We can also obtain the initial distribution for the case the focal player applies $\mathbf{q}$ analogously. Solving Eq.~\ref{equ:vbellman}, we can get the explicit form of the mean distribution $\mathbf{v}$
\begin{equation}
  \mathbf{v} =(1-\delta) \mathbf{v}(0)[\mathbf{I}-\delta \mathbf{M}']^{-1},
  \label{equ:v}
\end{equation}
where $\mathbf{I}$ is the identity matrix with the same dimension as $\mathbf{M}'$ and $(\cdot)^{-1}$ represents the inverse of a matrix. Substituting the results of Eq.~\ref{equ:v} to Eq.~\ref{equ:payoff}, we get long-term payoffs in finitely repeated multiplayer games.

\paragraph{Cooperation rate of memory-1 strategies.} Another application of reduced transition matrices is to improve the efficiency of computing the cooperation rates of strategies. The cooperation rate of a memory-1 strategy $\mathbf{p}$ is defined as the average cooperation rate of the group where all players apply $\mathbf{p}$. Given the mean distribution of outcomes, the cooperation rate is written as
\begin{equation}
  \eta_{\mathbf{p}}= \sum_{x} v_{C C_{x}^{\mathbf{p}}} \frac{x+1}{n} + \sum_{x} v_{D C_{x}^{\mathbf{p}}} \frac{x}{n}.
  \label{equ:cooprate}
\end{equation}
The dimension of $\mathbf{M}'$ is $2n$, which increases linearly with the number of players.

\section*{Acknowledgments}
We gratefully acknowledge the support from the National Natural Science Foundation of China (NSFC 62036002) and PKU-Baidu Fund (2020BD017).

\bibliographystyle{unsrt}  
\bibliography{references}

\begin{thebibliography}{10}

\bibitem{nowak2006five}
Martin~A. Nowak.
\newblock Five rules for the evolution of cooperation.
\newblock {\em Science}, 314(5805):1560--1563, 2006.

\bibitem{kennedy2005don}
Donald Kennedy and Colin Norman.
\newblock What don't we know?
\newblock {\em Science}, 309(5731):75--75, 2005.

\bibitem{nowak2006evolutionary}
Martin~A. Nowak.
\newblock {\em Evolutionary dynamics: exploring the equations of life}.
\newblock Harvard University Press, 2006.

\bibitem{su2019evolutionary}
Qi~Su, Alex McAvoy, Long Wang, and Martin~A. Nowak.
\newblock {Evolutionary dynamics with game transitions}.
\newblock {\em Proc. Natl. Acad. Sci. U.S.A.}, 116(51):25398--25404, 2019.

\bibitem{li2020evolution}
Aming Li, Lei Zhou, Qi~Su, Sean~P. Cornelius, Yang-Yu Liu, Long Wang, and
  Simon~A Levin.
\newblock Evolution of cooperation on temporal networks.
\newblock {\em Nat. Commun.}, 11(1):2259, 2020.

\bibitem{zhou2021aspiration}
Lei Zhou, Bin Wu, Jinming Du, and Long Wang.
\newblock Aspiration dynamics generate robust predictions in heterogeneous
  populations.
\newblock {\em Nat. Commun.}, 12(1):3250, 2021.

\bibitem{nowak1998evolution}
Martin~A. Nowak and Karl Sigmund.
\newblock Evolution of indirect reciprocity by image scoring.
\newblock {\em Nature}, 393(6685):573--577, 1998.

\bibitem{hauert2002volunteering}
Christoph Hauert, Silvia De~Monte, Josef Hofbauer, and Karl Sigmund.
\newblock Volunteering as red queen mechanism for cooperation in public goods
  games.
\newblock {\em Science}, 296(5570):1129--1132, 2002.

\bibitem{henrich2006costly}
Joseph Henrich, Richard McElreath, Abigail Barr, Jean Ensminger, Clark Barrett,
  Alexander Bolyanatz, Juan~Camilo Cardaroas, Michael Gurven, Edwins Gwako,
  Natalie Henrich, Carolyn Lesoronol, Frank Marlowe, David Tracer, and John
  Ziker.
\newblock {Costly punishment across human societies}.
\newblock {\em Science}, 312(5781):1767--1770, 2006.

\bibitem{nowak1993strategy}
Martin~A. Nowak and Karl Sigmund.
\newblock A strategy of win-stay, lose-shift that outperforms tit-for-tat in
  the prisoner's dilemma game.
\newblock {\em Nature}, 364(6432):56--58, 1993.

\bibitem{trivers1971evolution}
Robert~L. Trivers.
\newblock The evolution of reciprocal altruism.
\newblock {\em Q. Rev. Biol.}, 46(1):35--57, 1971.

\bibitem{pinheiro2014evolution}
Flavio~L. Pinheiro, Vitor~V. Vasconcelos, Francisco~C. Santos, and Jorge~M.
  Pacheco.
\newblock Evolution of all-or-none strategies in repeated public goods
  dilemmas.
\newblock {\em PLoS Comput. Biol.}, 10(11):e1003945, 2014.

\bibitem{press2012iterated}
William~H. Press and Freeman~J. Dyson.
\newblock Iterated prisoner's dilemma contains strategies that dominate any
  evolutionary opponent.
\newblock {\em Proc. Natl. Acad. Sci. U.S.A.}, 109(26):10409--10413, 2012.

\bibitem{Nowak2005}
Martin~A. Nowak and Karl Sigmund.
\newblock {Evolution of indirect reciprocity}.
\newblock {\em Nature}, 437(7063):1291--1298, 2005.

\bibitem{Hilbe2018}
Christian Hilbe, Laura Schmid, Josef Tkadlec, Krishnendu Chatterjee, and
  Martin~A. Nowak.
\newblock {Indirect reciprocity with private, noisy, and incomplete
  information}.
\newblock {\em Proc. Natl. Acad. Sci. U.S.A.}, 115(48):12241--12246, 2018.

\bibitem{Nowak1992}
Martin~A. Nowak and Robert~M. May.
\newblock {Evolutionary games and spatial chaos}.
\newblock {\em Nature}, 359(6398):826--829, 1992.

\bibitem{Ohtsuki2006}
Hisashi Ohtsuki, Christoph Hauert, Erez Lieberman, and Martin~A. Nowak.
\newblock {A simple rule for the evolution of cooperation on graphs and social
  networks}.
\newblock {\em Nature}, 441(7092):502--505, 2006.

\bibitem{hauert1997effects}
Christoph Hauert and Heinz~Georg Schuster.
\newblock Effects of increasing the number of players and memory size in the
  iterated prisoner's dilemma: a numerical approach.
\newblock {\em Proc. R. Soc. B}, 264(1381):513--519, 1997.

\bibitem{hilbe2015evolutionary}
Christian Hilbe, Bin Wu, Arne Traulsen, and Martin~A. Nowak.
\newblock Evolutionary performance of zero-determinant strategies in
  multiplayer games.
\newblock {\em J. Theor. Biol.}, 374(7):115--124, 2015.

\bibitem{yang2013nonlinear}
Wu~Yang, Wei Liu, Andr{\'e}s Vi{\~n}a, Mao-Ning Tuanmu, Guangming He, Thomas
  Dietz, and Jianguo Liu.
\newblock Nonlinear effects of group size on collective action and resource
  outcomes.
\newblock {\em Proc. Natl. Acad. Sci. U.S.A.}, 110(27):10916--10921, 2013.

\bibitem{hilbe2014cooperation}
Christian Hilbe, Bin Wu, Arne Traulsen, and Martin~A. Nowak.
\newblock Cooperation and control in multiplayer social dilemmas.
\newblock {\em Proc. Natl. Acad. Sci. U.S.A.}, 111(46):16425--16430, 2014.

\bibitem{wu2018coevolutionary}
Te~Wu, Feng Fu, and Long Wang.
\newblock Coevolutionary dynamics of aspiration and strategy in spatial
  repeated public goods games.
\newblock {\em New J. Phys.}, 20(6):063007, 2018.

\bibitem{wu2019phenotype}
Te~Wu, Feng Fu, and Long Wang.
\newblock Phenotype affinity mediated interactions can facilitate the evolution
  of cooperation.
\newblock {\em J. Theor. Biol.}, 462:361--369, 2019.

\bibitem{nowak2004emergence}
Martin~A. Nowak, Akira Sasaki, Christine Taylor, and Drew Fudenberg.
\newblock Emergence of cooperation and evolutionary stability in finite
  populations.
\newblock {\em Nature}, 428(6983):646--650, 2004.

\bibitem{imhof2010stochastic}
Lorens~A. Imhof and Martin~A. Nowak.
\newblock Stochastic evolutionary dynamics of direct reciprocity.
\newblock {\em Proc. R. Soc. B}, 277(1680):463--468, 2010.

\bibitem{stewart2013extortion}
Alexander~J. Stewart and Joshua~B. Plotkin.
\newblock From extortion to generosity, evolution in the iterated prisoner’s
  dilemma.
\newblock {\em Proc. Natl. Acad. Sci. U.S.A.}, 110(38):15348--15353, 2013.

\bibitem{hilbe2017memory}
Christian Hilbe, Luis~A Martinez-Vaquero, Krishnendu Chatterjee, and Martin~A.
  Nowak.
\newblock Memory-n strategies of direct reciprocity.
\newblock {\em Proc. Natl. Acad. Sci. U.S.A.}, 114(18):4715--4720, 2017.

\bibitem{pan2015zero}
Liming Pan, Dong Hao, Zhihai Rong, and Tao Zhou.
\newblock Zero-determinant strategies in iterated public goods game.
\newblock {\em Sci. Rep.}, 5(1):13096, 2015.

\bibitem{boyd1988evolution}
Robert Boyd and Peter~J. Richerson.
\newblock The evolution of reciprocity in sizable groups.
\newblock {\em J. Theor. Biol.}, 132(3):337--356, 1988.

\end{thebibliography}


\newpage
\begin{thebibliography}{10}

\bibitem{nowak1993strategy}
Nowak, M. A. and Sigmund, K.
  \newblock A strategy of win-stay, lose-shift that outperforms tit-for-tat in the Prisoner's Dilemma game.
  \newblock {\em Nature.} 364(6432):56-58, 1993.
  
  \bibitem{bellman1957markovian}
  Bellman, R.
  \newblock A Markovian decision process.
  \newblock {\em J. Math. Mech.} 6(5):679--684, 1957.
  
\end{thebibliography}

\pagebreak

\begin{spacing}{1.5}
\begin{center}
  \textbf{\Large Supplementary Information for  \\ 
State-clustering method of payoff computation in repeated multiplayer games}\\[.2cm]
  Fang Chen, Te Wu, Guocheng Wang and Long Wang\\[.1cm]
\end{center}

\setcounter{equation}{0}
\setcounter{figure}{0}
\setcounter{table}{0}
\setcounter{page}{1}
\setcounter{section}{0}
\renewcommand{\theequation}{S\arabic{equation}}
\renewcommand{\thefigure}{S\arabic{figure}}

In the following, we provide detailed derivations and proofs of our method to calculate the long-term payoffs for both infinitely and finitely repeated multiplayer games. We prove that our method is not confined to the assumption of low mutations but still works in the repeated games where more than two strategies are available. Moreover, we report further results to show the robustness of our findings. In Section \ref{sec:Smodel}, we give a full description of repeated multiplayer games and memory-1 strategies. Section \ref{sec:payoff} provides the details of calculating the long-term payoffs with two or more strategies. Finally, the appendix contains the proofs of all theorems.

\section{Repeated multiplayer games and memory-1 strategies.}
\label{sec:Smodel}
Consider a repeated multiplayer game with a group of $n \geq 2$ players. In each round, players can choose an action from their action set $\mathcal{A}=\{C,D\}$. Here, $C$ represents cooperation and $D$ means defection. Assume the multiplayer game is symmetrical such that a player's payoff only depends on his action and the number of cooperators among his co-players. If there are $j$ cooperators among other players, a cooperator gets $a_j$ and a defector obtains $b_j$ (fig.~\ref{Sfig:model}). In this paper, we pay attention to the evolution of cooperation in multiplayer social dilemma, for which the following three conditions are necessary:
\begin{itemize}
  \item [i)] players always prefer their co-players to cooperate, i.e. $a_{j+1}\geq a_j$ and $b_{j+1} \geq b_j$,
  \item [ii)] a defector gets strictly more than a cooperator in any cases, i.e. $b_{j+1}>a_j$,
  \item [iii)] mutual cooperation is favoured over mutual defection, i.e. $a_{n-1}>b_0$. 
\end{itemize}
Public Goods Game is a special case of multiplayer social dilemma. The action of contributing an endowment $c>0$ to a common pool is regarded as cooperation. Total contributions in the common pool are multiplied by a synergy factor $r$ with $1<r<n$ and then evenly divided to each player. Thus, a cooperator's payoff is $a_j=(j+1)rc/n-c$ and a defector's is $b_j=jrc/n$ when there are $j$ cooperators among the rest. 

\begin{table}[ht]
  \centering
    \caption{\textbf{Payoffs of multiplayer games.} Consider a symmetrical $n$-player game in which each player either cooperates or defects. The payoff of each player depends on his own action and the number of cooperators among his co-players. If $j$ co-players cooperate, a cooperator will get $a_j$ and a defector will obtain $b_j$.}
  \begin{tabular}{c|cccccc}
    \hline
         \textbf{number of cooperators among co-players} & 0 & 1 & $\cdots$ & $j$ & $\cdots$ & $n-1$  \\ \hline
         cooperator's payoff & $a_0$ & $a_1$ & $\cdots$ & $a_j$ & $\cdots$ & $a_{n-1}$  \\ \hline
         defector's payoff & $b_0$ & $b_1$ & $\cdots$ & $b_j$ & $\cdots$ & $b_{n-1}$ \\
         \hline
    \end{tabular}

  \label{Sfig:model}
\end{table}

When a round of game terminates, another round occurs with a constant probability $\delta>0$. If $0<\delta<1$, the game will be played for finite rounds with the average rounds $1/(1-\delta)$. For $\delta=1$, the game will proceed infinitely. Let $h=(A_1,...,A_n)\in\mathcal{A}^n$ denote the state which contains all players' actions in a round. $\mathcal{H}$ denotes the set of all possible states. A \textit{memory-1} strategy tells players how to act in the next round according to the state in the previous round. For a symmetrical repeated multiplayer social dilemma, a memory-1 strategy can be written as a vector
\begin{equation}
  \mathbf{p}=(p_0,p_{C0},...,p_{Cn-1},p_{D0},...,p_{Dn-1}),
  \label{equ:strategy}
\end{equation} 
where $p_{Ak}$ is the probability that the focal player cooperates if his action is $A$ and $k$ co-players cooperate in the previous round. $p_0$ is the initial cooperation probability of the focal player in the first round. Suppose players do not execute their actions perfectly but are subject to a ``trembling hand'': when a player means to cooperate (defect), he defects (cooperates) by mistakes with a probability $\epsilon$. For a player with strategy $\mathbf{p}$, his effective strategy becomes $(1-\epsilon)\mathbf{p}+\epsilon(1-\mathbf{p})$. Such assumption guarantees all states are accessible, which is necessary when computing the expected payoffs during the iterated games.

\section{Long-term payoffs}\label{sec:payoff}
Given all players' strategies, long-term payoffs during iterated interactions can be calculated explicitly. The repeated game can be modeled as a Markov chain. We use $\mathbf{p}^i=(p_0^i,p_{C0}^i,...,p_{Cn-1}^i,p_{D0}^i,...,p_{Dn-1}^i)$ to denote the strategy of player $i$ and $\sigma$ to denote the total number of cooperators in state $h$. The transition probability from $h=(A_1,...,A_n)$ to $h'=(A'_1,...,A'_n)$ is given by product
\begin{equation}
  m_{h,h'}=\prod_{i=1}^{n} u_{i},
  \label{equ:M}
\end{equation}
where
\begin{equation*}
  u_{i}=\left\{\begin{array}{ll}
    (1-\epsilon) p_{C \sigma-1}^i+\epsilon\left(1-p_{C \sigma-1}^i\right), & \text { if } A_{i}=C \text { and } A_{i}^{\prime}=C \\
    (1-\epsilon) p_{D \sigma}^i+\epsilon\left(1-p_{D \sigma}^i\right), & \text { if } A_{i}=D \text { and } A_{i}^{\prime}=C \\
    \epsilon p_{C \sigma-1}^i+(1-\epsilon)\left(1-p_{C \sigma-1}^i\right), & \text { if } A_{i}=C \text { and } A_{i}^{\prime}=D \\
    \epsilon p_{D \sigma}^i+(1-\epsilon)\left(1-p_{D \sigma}^i\right), & \text { if } A_{i}=D \text { and } A_{i}^{\prime}=D
    \end{array}\right..
\end{equation*}
Let $\mathbf{v}(t)=(v_h(t))_h$ denote the distribution vector of all states at time $t$. Collecting all probabilities in Eq.~\ref{equ:M}, we get the $2^n \times 2^n$ transition matrix $\mathbf{M} = (m_{h,h'})$. The distribution of states at time $t+1$ is $v(t+1)=v(t)\mathbf{M}$. For finitely repeated game, the mean distribution of $\mathbf{v}(t)$, denoted by $\mathbf{v}$, is given by
\begin{equation}
  \sum_{t=0}^{\infty}(1-\delta) \delta^{t} \mathbf{v}(t),
  \label{definition:vfiinite}
\end{equation}
where $(1-\delta)\delta^t$ can be interpreted as the probability that the game proceeds for $n$ rounds. According to Bellman equation \cite{bellman1957markovian}, Eq.~\ref{definition:vfiinite} becomes
\begin{equation}
  \begin{aligned}
    \mathbf{v} &=\sum_{t=0}^{\infty}(1-\delta) \delta^{t} \mathbf{v}(t) \\
    &=(1-\delta) \mathbf{v}(0)+\sum_{t=1}^{\infty}(1-\delta) \delta^{t} \mathbf{v}(t) \\
    &=(1-\delta) \mathbf{v}(0)+\sum_{t=0}^{\infty}(1-\delta) \delta^{t+1} \mathbf{v}(t+1) \\
    &=(1-\delta) \mathbf{v}(0)+\delta \sum_{t=0}^{\infty}(1-\delta) \delta^{t} \mathbf{v}(t) \mathbf{M} \\
    &=(1-\delta) \mathbf{v}(0)+\delta \mathbf{v} \mathbf{M},
  \end{aligned}
  \label{computevfinite}
\end{equation}
where $\mathbf{v}(0)=\prod_{i=1}^n p_0^i$ is the initial distribution of states. Solving Eq.~\ref{computevfinite}, we obtain the mean distribution $\mathbf{v}(t)$ as
\begin{equation}
  \mathbf{v} =(1-\delta) \mathbf{v}(0)[\mathbf{I}-\delta \mathbf{M}]^{-1},
  \label{sequ:v}
\end{equation}
where $\mathbf{I}$ is the identity matrix with the same dimension as $\mathbf{M}$. If the game proceeds for infinite rounds, the mean distribution of $\mathbf{v}(t)$ is the unique left eigenvector of the transition matrix $\mathbf{M}$, that is,
\begin{equation}
  \mathbf{v}=\mathbf{v}\mathbf{M}.
  \label{definition:vinfinite}
\end{equation}
Therefore, the expected long-term payoff of player $i$ is given by
\begin{equation}
  \pi_i = \sum_{h \in \mathcal{H}} v_hs_h,
  \label{sequ:payoff}
\end{equation}
where $s_h$ is the payoff of player $i$ in $h$.

Given the mean distribution $\mathbf{v}$, we can also compute the cooperation rate of a strategy. The cooperation rate of a strategy is defined as the cooperation rate if all players in the group  apply this strategy. Thus, the cooperation rate of a strategy is
\begin{equation}
  \eta_{\mathbf{p}} = \sum_{h \in \mathcal{H}} \frac{\sigma}{n} v_h.
  \label{definition:cooperationrate}
\end{equation}

Given the transition matrix, it seems not difficult to compute the players' payoffs. However, note that the dimension of $\mathbf{M}$ increases exponentially as the group size $n$ grows. For a repeated game with 20 players, it is impossible for a personal computer to calculate long-term payoffs since we need to solve the inverse of a square matrix of order $2^{20} = 1048576$. It's also difficult to study the evolution of cooperation in a group with 10 players as the process of solving inverse is repeated for numerous times. Here, we propose a state-clustering technique to reduce the dimension of the transition matrix. In the following, we shall introduce detailed derivations of the reduced transition matrix with and without the limit of low mutations.
\subsection{State-clustering technique with the limit of low mutations}
Let's begin with the case that the mutations are rare, where a population contains at most two different strategies at the same time. Let $\mathbf{p}$ and $\mathbf{q}$ denote the two strategies. $A C_x^{\mathbf{p}} C_y^{\mathbf{q}}$ denotes the state where the focal player takes the action $A$, $x$ players cooperate among players applying $\mathbf{p}$ and $y$ players cooperate among those adopting $\mathbf{q}$. For convenience, we introduce a map $f: \mathcal{A} \rightarrow \mathbb{R}$ with $f(C)=1$ and $f(D)=0$. If the state in the previous round is $A C_x^{\mathbf{p}} C_y^{\mathbf{q}}$, the focal player with $\mathbf{p}$ ($\mathbf{q}$) cooperates with probability $p_{Ax+y}$ ($q_{Ax+y}$). 
The $\mathbf{p}$-players who cooperate (defect) in the previous round will cooperate in the next round with probability $p_{Cx+y-1+f(A)}$ ($p_{Dx+y+f(A)}$). Similarly, the $\mathbf{q}$-players who cooperate (defect) in the previous round will cooperate with probability $q_{Cx+y-1+f(A)}$ ($q_{Dx+y+f(A)}$). Suppose $k$ players apply $\mathbf{p}$ and $n-1-k$ players adopt $\mathbf{q}$ apart from the focal player. Let $A C_x^{\mathbf{p}} C_y^{\mathbf{q}}$ where $A \in \mathcal{A}$, $x \in \{0,1,..,k\}$ and $y \in \{0,1,...,n-1-k\}$ represent the state of the Markov chain. The transition probability from $A C_x^{\mathbf{p}} C_y^{\mathbf{q}}$ to $A' C_{x'}^{\mathbf{p}} C_{y'}^{\mathbf{q}}$ is given by
\begin{equation}
  \begin{array}{ll}
    \left|1-f(A')-p_{Ax+y}\right| 
    &\left[\sum_{j=0}^{x^{\prime}} \tbinom{x}{j}p_{Cx+y-1+f(A)}^{j}\left(1-p_{Cx+y-1+f(A)}\right)^{x-j}\right. \\
    &\left.\tbinom{k-x}{x'-j} p_{Dx+y+f(A)}^{x^{\prime}-j}\left(1-p_{Dx+y+f(A)}\right)^{(k-x)-\left(x^{\prime}-j\right)}\right] \\
    &{\left[\sum_{j=0}^{y^{\prime}}\tbinom{y}{j} q_{Cx+y-1+f(A)}^{j}\left(1-q_{Cx+y-1+f(A)}\right)^{y-j}\right.} \\
    &\left.\tbinom{n-k-1-y}{y'-j} q_{Dx+y+f(A)}^{y^{\prime}-j}\left(1-q_{Dx+y+f(A)}\right)^{(n-1-k-y)-\left(y^{\prime}-j\right)}\right]
    \end{array}
    \label{equ:transitionProbability}
\end{equation}
for the focal player with $\mathbf{p}$. We can also get the transition probability by substituting $p_{Ax+y}$ by $q_{Ax+y}$ if the focal player adopts $\mathbf{q}$. Here, $|a|$ is the absolute value of $a$. $j$ represents the number of players who cooperate in both present and previous rounds. Collect all probabilities in Eq.~\ref{equ:transitionProbability} into a transition matrix $\mathbf{M}'$ of order $2(k+1)(n-k)$. The dimension of the reduced transition matrix $\mathbf{M}'$ reaches the maximum $2(\lfloor \frac{n-1}{2} \rfloor +1)(n-\lfloor \frac{n-1}{2} \rfloor)$ at $k=\lfloor\frac{n-1}{2}\rfloor$, which increases with the group size at a square rate rather than an exponential rate. Let $\mu_{A C_x^{\mathbf{p}} C_y^{\mathbf{q}}}$ denote the average frequency that states $A C_x^{\mathbf{p}} C_y^{\mathbf{q}}$ occurs during iterated interactions. In finitely repeated multiplayer game, the mean distribution $\mathbb{\mu}$ is
$$ \mathbf{\mu} =(1-\delta) \mathbf{\mu}(0)[\mathbf{I}-\delta \mathbf{M}']^{-1}, $$
according to the Bellman equation. Here $\mathbf{\mu}(0)$ is the initial distribution of $A C_x^{\mathbf{p}} C_y^{\mathbf{q}}$ and each element can be written as $$\mu_{A C_x^{\mathbf{p}} C_y^{\mathbf{q}}}(0)= \left|1-f(A)-p_{0}\right|\tbinom{k}{x} p_{0}^{x}\left(1-p_{0}\right)^{k-x}\tbinom{n-k-1}{y} q_{0}^{y}\left(1-q_{0}\right)^{n-1-k-y},$$ for the focal player with $\mathbf{p}$. For infinitely multiplayer game, $\mathbf{\mu}$ is given by the unique left eigenvector of $\mathbf{M}'$. Let $\mathcal{A}_{Axy}\subseteq \mathcal{H}$ denote the set of the Markov chain $\mathbf{M}$'s states where the focal player takes action $A$, $x$ players cooperate among players applying $\mathbf{p}$ and $y$ players cooperate among those with $\mathbf{q}$. We prove the $\mathbf{v}$ and $\mathbf{\mu}$ are related as follows.
\newtheorem{theorem}{Theorem}
\begin{theorem}
  For a repeated symmetrical game as shown in Tab.~\ref{Sfig:model}, the mean distribution of state $h$ and the mean distribution of states $A C_x^{\mathbf{p}} C_y^{\mathbf{q}}$ are related as follows:
  \begin{equation}
    \mu_{A C_x^{\mathbf{p}} C_y^{\mathbf{q}}}=\sum_{h \in \mathcal{A}_{Axy}}v_h,
    \label{equ:relationship}
  \end{equation}
  for any $A\in\{C,D\}$, $x\in\{0,...,k\}$ and $y\in\{0,...,n-1-k\}$ in both infinitely and finitely repeated games.
  \label{the:1}  
\end{theorem}

Given the mean distribution $\mathbf{\mu}$, we can also calculate the payoff of each player and cooperation rates of strategies. The long-term payoffs of player $i$ is given by
$$\pi_i = \sum_{x,y} \mu_{C C_{x}^{\mathbf{p}} C_y^{\mathbf{q}}} a_{x+y} + \sum_{x,y} \mu_{D C_{x}^{\mathbf{p}} C_y^{\mathbf{q}}} b_{x+y}.$$
The cooperation rate of strategy $\mathbf{p}$ can be obtained by
$$\sum_{x,y} \mu_{C C_{x}^{\mathbf{p}}} \frac{x+1}{n} + \sum_{x,y} \mu_{D C_{x}^{\mathbf{p}} } \frac{x}{n}.$$
Therefore, the following two corollaries hold.
\newtheorem{corollary}{Corollary}
\begin{corollary}
  For a repeated symmetrical game as shown in Tab.~\ref{Sfig:model}, the long-term payoffs calculated by transition matrix $\mathbf{M}$ and reduced transition matrix $\mathbf{M}'$ are the same under the limit of low mutations.
\end{corollary}

\newtheorem{corollary1}{Corollary}
\begin{corollary}
  For a repeated symmetrical game as shown in Tab.~\ref{Sfig:model}, the cooperation rates of a strategy calculated by transition matrix $\mathbf{M}$ and reduced transition matrix $\mathbf{M}'$ are the same under the limit of low mutations.
\end{corollary}
\subsection{State-clustering technique without the limit of rare mutations}
Next, let's calculate the reduced transition matrix without the limit of rare mutations. Suppose there are $m$ strategies in the population and let $\mathbf{p}^{(i)}=(p^{(i)}_{C0},...,p^{(i)}_{Cn-1},p^{(i)}_{D0},...,p^{(i)}_{Dn-1}), i \in \{1,2,...,m\}$ denote these strategies. We can group the states of the Markov chain into $A C_{x_1}^{\mathbf{p}^{(1)}} \cdot \cdot \cdot C_{x_m}^{\mathbf{p}^{(m)}}$, where $A$ is the action of the focal player and $C_{x_i}^{\mathbf{p}^{(i)}}$ represents $x_i$ players cooperate among those with $\mathbf{p}^{(i)}$. Suppose there are $k_i$ players applying strategy $\mathbf{p}^{(i)}$ and $\sum_{i=1}^m k_i =n-1$. The transition probability from the state $A C_{x_1}^{\mathbf{p}^{(1)}} \cdots C_{x_m}^{\mathbf{p}^{(m)}}$ to the state $A' C_{x'_1}^{\mathbf{p}^{(1)}} \cdot \cdot \cdot C_{x'_m}^{\mathbf{p}^{(m)}}$ is
\begin{equation}
  \begin{aligned}
    \left|1-f(A')-p_{Ax+y} ^{(i)}\right| \prod_{i=1}^m 
    &\sum_{j=0}^{x^{\prime}_i}\tbinom{x_i}{j}(p_{Cx+y-1+f(A)}^{(i)})^{j}\left(1-p_{Cx+y-1+f(A)}^{(i)}\right)^{x_i-j}\\
    &\tbinom{k_i-x_i}{x'_i-j}(p_{Dx+y+f(A)}^{(i)})^{x'_i-j}\left(1-p_{Dx+y+f(A)}^{(i)}\right)^{(k_i-x_i)-\left(x^{\prime}_i-j\right)}
  \end{aligned}
  \label{equ:transitionProbabilitym}
\end{equation}
for the focal player with $\mathbf{p}^{(i)}$. Collect all probabilities in Eq.~\ref{equ:transitionProbabilitym} into a transition matrix $\mathbf{M}''$ of order $2\prod_{i=1}^m (k_i+1)$. The number of reduced states is less than that of $\mathbf{M}$ unless $m \ge n$. Thus, our method can also reduce the dimension of transition matrix even when there are multiple strategies.

Let $\nu$ denote the mean distribution of $A C_{x_1}^{\mathbf{p}^{(1)}} \cdot \cdot \cdot C_{x_m}^{\mathbf{p}^{(m)}}$. $\nu$ is given by
$$\mathbf{\nu} =(1-\delta) \mathbf{\nu}(0)[\mathbf{I}-\delta \mathbf{M}'']^{-1}.$$
in the finitely repeated games. $\nu$ satisfies
$$\mathbf{\nu} = \mathbf{\nu}\mathbf{M}''$$
in the infinitely repeated games.
Here, $\mathbf{I}$ is the identity matrix of order $2\prod_{i=1}^m (k_i+1)$. $\mathbf{\nu}(0)$ is the initial distribution of $A C_{x_1}^{\mathbf{p}^{(1)}} \cdot \cdot \cdot C_{x_m}^{\mathbf{p}^{(m)}}$ and each element is
$$\nu_{A C_{x_1}^{\mathbf{p}^{(1)}} \cdot \cdot \cdot C_{x_m}^{\mathbf{p}^{(m)}}}(0)=\left|1-f(A)-p_0^{(i)}\right|\prod_{i=1}^m \tbinom{k_i}{x_i} (p_{0}^{(i)})^{x_i}\left(1-p_{0}^{(i)}\right)^{k_i-x_i}.$$
Let $\mathcal{A}_{Ax_1 \cdots x_m}$ denote the set of the states where the focal player takes the action $A$ and $x_i$ players cooperate among those applying $\mathbf{p}^{(i)}$. The following relationship between $\mathbf{\nu}$ and $\mathbf{v}$ holds.
\newtheorem{theorem2}{Theorem}
\begin{theorem}
  For a repeated symmetrical game as shown in Tab.~\ref{Sfig:model}, the mean distribution of states $h$ and the mean distribution of states $A C_{x_1}^{\mathbf{p}^{(1)}} \cdot \cdot \cdot C_{x_m}^{\mathbf{p}^{(m)}}$ are related as follows:
  \begin{equation}
    \nu_{A C_{x_1}^{\mathbf{p}^{(1)}} \cdots C_{x_m}^{\mathbf{p}^{(m)}}}=\sum_{h \in \mathcal{A}_{Ax_1 \cdots x_m}}v_h,
    \label{equ:relationshipm}
  \end{equation}
  for any $A$ and $x_i\in\{0,1,...,k_i\}$ in both infinitely and finitely repeated games.
  \label{the:2}  
\end{theorem}

Given the mean distribution $\mathbf{\nu}$, the long-term payoffs of player $i$ can be written as
$$\pi_i = \sum_{x_1, \cdots, x_m} \left(\nu_{C C_{x_1}^{\mathbf{p}^{(1)}} \cdots C_{x_m}^{\mathbf{p}^{(m)}}} a_{\sum_{j=1}^mx_j} + \nu_{D C_{x_1}^{\mathbf{p}^{(1)}} \cdots C_{x_m}^{\mathbf{p}^{(m)}}} b_{\sum_{j=1}^mx_j}\right).$$
Thus, we have the following  corollary.
\newtheorem{corollary2}{Corollary}
\begin{corollary}
  For a repeated symmetrical game as shown in Tab.~\ref{Sfig:model}, the long-term payoffs calculated by transition matrix $\mathbf{M}$ and reduced transition matrix $\mathbf{M}''$ are the same if there are $m$ strategies available in the population. 
\end{corollary}

\section*{Appendix}
\paragraph{proof of theorem \ref{the:1}}:
\newtheorem{proof1}{Proof}
\begin{proof}
We use Mathematical Induction (MI) to prove theorem \ref{the:1}. To do this, we first prove $\mu_{A C_x^{\mathbf{p}} C_y^{\mathbf{q}}}(0)=\sum_{h \in \mathcal{A}_{Axy}}v_h(0)$ for any $A\in\{C,D\}$, $x\in\{0,...,k\}$ and $y\in\{0,...,n-1-k\}$. Then, we prove $\mu_{A C_x^{\mathbf{p}} C_y^{\mathbf{q}}}(t+1)=\sum_{h \in \mathcal{A}_{Axy}}v_h(t+1)$ holds for any $A$, $x$ and $y$ if $\mu_{A C_x^{\mathbf{p}} C_y^{\mathbf{q}}}(t)=\sum_{h \in \mathcal{A}_{Axy}}v_h(t)$ holds for any $A$, $x$ and $y$. Finally, we prove $\mu_{A C_x^{\mathbf{p}} C_y^{\mathbf{q}}}=\sum_{h \in \mathcal{A}_{Axy}}v_h$ for any $A$, $x$ and $y$ in both infinitely and finitely repeated games. Here, we only discuss the case where the focal player adopts $\mathbf{p}$. The other case in which the focal player applies $\mathbf{q}$ can be proved similarly. 

When $t=0$, for any state $h \in \mathcal{A}_{Axy}$, it occurs with probability $\left|1-f(A)-p_0\right|p_{0}^{x}(1-p_{0})^{k-x} q_{0}^{y}(1-q_{0})^{n-1-k-y}$. There are $\tbinom{k}{x}$ states where $x$ cooperators adopt $\mathbf{p}$ and $\tbinom{n-1-k}{y}$ states where $y$ cooperators adopt $\mathbf{q}$. Thus, $\sum_{h \in \mathcal{A}_{Axy}} v_h(0) = \tbinom{k}{x}\tbinom{n-1-k}{y}\left|1-f(A)-p_0\right|p_{0}^{x}(1-p_{0})^{k-x} q_{0}^{y}(1-q_{0})^{n-1-k-y}$, which equals to the probability that $A C_x^{\mathbf{p}} C_y^{\mathbf{q}}$ occurs. There are no restrictions on $A$, $x$ and $y$, leading to that $\mu_{A C_x^{\mathbf{p}} C_y^{\mathbf{q}}}(0)=\sum_{h \in \mathcal{A}_{Axy}}v_h(0)$ holds for any $A$, $x$ and $y$.

Now, we suppose 
\begin{equation}
  \mu_{A C_x^{\mathbf{p}} C_y^{\mathbf{q}}}(t)=\sum_{h \in \mathcal{A}_{Axy}}v_h(t).
  \label{equ:vt}
\end{equation}
for any $A\in\{C,D\}$, $x\in\{0,...,k\}$ and $y\in\{0,...,n-1-k\}$. We shall prove Eq.~\ref{equ:vt} still holds at $t+1$ in the following. To do this, we first prove the sum of transition probabilities from each state in $\mathcal{A}_{Axy}$ to all states in $\mathcal{A}_{A'x'y'}$ is equal to the transition probability $m'_{A C_x^{\mathbf{p}} C_y^{\mathbf{q}},A' C_{x'}^{\mathbf{p}} C_{y'}^{\mathbf{q}}}$. Select a random state $\tilde{h}$ from $\mathcal{A}_{Axy}$.
In any state $\hat{h} \in \mathcal{A}_{A'x'y'}$, each cooperator either cooperates or defects in the previous round. If $j$ players with $\mathbf{p}$ cooperates in $\tilde{h}$, the transition probability from $x$ $\mathbf{p}$ players cooperating to $x'$ $\mathbf{p}$ players cooperating is $$\tbinom{x}{j}p_{Cx+y-1+f(A)}^j(1-p_{Cx+y-1+f(A)})^{x-j} \tbinom{k-x}{x'-j}p_{Dx+y+f(A)}^{x'-j}(1-p_{Dx+y+f(A)})^{(k-x)-(x'-j)}.$$ If $j$ players with $\mathbf{q}$ cooperate in $\tilde{h}$, the transition probability from $y$ $\mathbf{q}$ players cooperating to $y'$ $\mathbf{q}$ players cooperating is $$\tbinom{y}{j}q_{Cx+y-1+f(A)}^j(1-q_{Cx+y-1+f(A)})^{y-j} \tbinom{n-k-1-y}{y'-j}q_{Dx+y+f(A)}^{y'-j}(1-q_{Dx+y+f(A)})^{(n-k-1-y)-(y'-j)}.$$ 
We define $\tbinom{x}{x'}=0$ for $x'>x$. The sum of transition probabilities from $\tilde{h}$ to all states in $\mathcal{A}_{A'x'y'}$ is 
$$\begin{array}{ll}
  \left|1-f(A')-p_{Ax+y}\right| &\left[\sum_{j=0}^{x^{\prime}} \tbinom{x}{j}p_{Cx+y-1+f(A)}^j(1-p_{Cx+y-1+f(A)})^{x-j}\right.\\
  &\left.\tbinom{k-x}{x'-j}p_{Dx+y+f(A)}^{x'-j}(1-p_{Dx+y+f(A)})^{(k-x)-(x'-j)} \right] \\
  &{\left[\sum_{j=0}^{y^{\prime}} \tbinom{y}{j} q_{Cx+y-1+f(A)}^{j}\left(1-q_{Cx+y-1+f(A)}\right)^{y-j}\right.} \\
  &\left. \tbinom{n-k-1-y}{y'-j} q_{Dx+y+f(A)}^{y^{\prime}-j}\left(1-q_{Dx+y+f(A)}\right)^{(n-1-k-y)-\left(y^{\prime}-j\right)}\right],
\end{array}$$
which equals to $m'_{A C_x^{\mathbf{p}} C_y^{\mathbf{q}},A' C_{x'}^{\mathbf{p}} C_{y'}^{\mathbf{q}}}$. That is, $ \sum_{h'\in \mathcal{A}_{A'x'y'}} m_{h,h'}=m'_{A C_x^{\mathbf{p}} C_y^{\mathbf{q}},A' C_{x'}^{\mathbf{p}} C_{y'}^{\mathbf{q}}}$.

Given the distribution of states at time $t$, $\mathbf{v}(t+1)=(v_h(t+1))_h$ can be written as
$$v_h(t+1) = \sum_{h'\in \mathcal{A}^n} v_{h'}(t)m_{h',h}.$$
Thus,
\begin{equation}
  \begin{aligned}
    \sum_{h\in\mathcal{A}_{Axy}} v_h(t+1) &= \sum_{h\in\mathcal{A}_{Axy}} \sum_{h'\in \mathcal{A}^n} v_{h'}(t)m_{h',h}\\
    &=\sum_{h\in\mathcal{A}_{Axy}}\sum_{A',x',y'} \sum_{h'\in\mathcal{A}_{A'x'y'}}v_{h'}(t)m_{h',h}\\
    &=\sum_{A',x',y'}\sum_{h'\in\mathcal{A}_{A'x'y'}}v_{h'}(t)\sum_{h\in\mathcal{A}_{Axy}}m_{h',h}\\
    &=\sum_{A',x',y'}\sum_{h'\in\mathcal{A}_{A'x'y'}}v_{h'}(t)m'_{A' C_{x'}^{\mathbf{p}} C_{y'}^{\mathbf{q}},A C_x^{\mathbf{p}} C_y^{\mathbf{q}}}\\
    &=\sum_{A',x',y'}\mu_{A' C_{x'}^{\mathbf{p}} C_{y'}^{\mathbf{q}}}(t)m'_{A C_{x'}^{\mathbf{p}} C_{y'}^{\mathbf{q}},A C_x^{\mathbf{p}} C_y^{\mathbf{q}}}\\
    &=\mu_{A C_x^{\mathbf{p}} C_y^{\mathbf{q}}}(t+1).
  \end{aligned}
\end{equation}

In summary, we have proved $\mu_{A C_x^{\mathbf{p}} C_y^{\mathbf{q}}}(t)=\sum_{h \in \mathcal{A}_{Axy}}v_h(t)$ at any time $t$. Finally, we prove Eq.~\ref{equ:relationship} holds in both infinitely and finitely repeated games. In finitely repeated multiplayer games, the mean distribution $\mu_{A C_x^{\mathbf{p}} C_y^{\mathbf{q}}}$ is give by
\begin{equation}
  \begin{aligned}
    \mu_{A C_x^{\mathbf{p}} C_y^{\mathbf{q}}} &= (1-\delta)\sum_{t=0}^{\infty} \delta^t \mu_{A C_x^{\mathbf{p}} C_y^{\mathbf{q}}} (t)\\
    &=(1-\delta) \sum_{t=0}^{\infty} \sum_{h \in \mathcal{A}_{Axy}}\delta^t v_h(t)\\
    &=(1-\delta) \sum_{h \in \mathcal{A}_{Axy}}\sum_{t=0}^{\infty} \delta^t v_h(t)\\
    &= \sum_{h \in \mathcal{A}_{Axy}} v_h.
  \end{aligned} 
  \label{equ:finitere} 
\end{equation}
In infinitely repeated games, $\mathbf{v}$ and $\mathbf{\mu}$ are the left eigenvector of $\mathbf{M}$ and $\mathbf{M}'$, respectively. For $\epsilon >0$ the matrix $\mathbf{M}$ and $\mathbf{M}'$ are stochastic and primitive. Hence, $\mathbf{v}=\lim_{t \rightarrow \infty}\mathbf{v}(t)$ and $\mathbf{\mu}_{A C_x^{\mathbf{p}} C_y^{\mathbf{q}}} = \lim_{t \rightarrow \infty} \mathbf{\mu}_{A C_x^{\mathbf{p}} C_y^{\mathbf{q}}}(t)$ hold. So,
\begin{equation}
  \mu_{A C_x^{\mathbf{p}} C_y^{\mathbf{q}}}=\sum_{h \in \mathcal{A}_{Axy}} v_h.
  \label{equ:infinitere}
\end{equation}
Therefore, Eq.~\ref{equ:relationship} holds. 

\end{proof}

\paragraph{proof of theorem~\ref{the:2}}:
\newtheorem{proof2}{Proof}
\begin{proof}
We use Mathematical Induction (MI) to prove the theorem \ref{the:2}. To do this, we first prove $\nu_{A C_{x_1}^{\mathbf{p}^{(1)}} \cdots C_{x_m}^{\mathbf{p}^{(m)}}}(0)=\sum_{h \in \mathcal{A}_{Ax_1 \cdots x_m}}v_h(0)$ for any $A\in\{C,D\}$ and $x_i\in\{0,...,k_i\}$. Then, we prove $\nu_{A C_{x_1}^{\mathbf{p}^{(1)}} \cdots C_{x_m}^{\mathbf{p}^{(m)}}}(t+1)=\sum_{h \in \mathcal{A}_{Ax_1 \cdots x_m}}v_h(t+1)$ holds for any $A$ and $x_i$ with $i\in\{1,...,m\}$ if $\nu_{A C_{x_1}^{\mathbf{p}^{(1)}} \cdots C_{x_m}^{\mathbf{p}^{(m)}}}(t)=\sum_{h \in \mathcal{A}_{Ax_1 \cdots x_m}}v_h(t)$ holds for any $A$ and $x_i$. Finally, we prove $\nu_{A C_{x_1}^{\mathbf{p}^{(1)}} \cdots C_{x_m}^{\mathbf{p}^{(m)}}}=\sum_{h \in \mathcal{A}_{Ax_1 \cdots x_m}}v_h$ for any $A$ and $x_i$ in both infinitely and finitely repeated games. 

When $t=0$, any state $h \in \mathcal{A}_{Ax_1 \cdots x_m}$ occurs with probability $$\left|1-f(A)-p_0^{(i)}\right|\prod_{i=1}^m(p_{0}^{(i)})^{x_i}\left(1-p_{0}^{(i)}\right)^{k_i-x_i}$$ according to Eq.~\ref{equ:M}. 
There are $\tbinom{k_i}{x_i}$ possible states if there are $x_i$ cooperators among $k_i$ $\mathbf{p}^{(i)}$-players. Thus, $$\sum_{h \in \mathcal{A}_{Ax_1 \cdots x_m}}v_h(0)=\left|1-f(A)-p_0^{(i)}\right|\prod_{i=1}^m \tbinom{k_i}{x_i} (p_{0}^{(i)})^{x_i}\left(1-p_{0}^{(i)}\right)^{k_i-x_i},$$ which also equals to the initial distribution $\nu_{A C_{x_1}^{\mathbf{p}^{(1)}} \cdots C_{x_m}^{\mathbf{p}^{(m)}}}(0)$.

Suppose 
\begin{equation}
  \nu_{A C_{x_1}^{\mathbf{p}^{(1)}} \cdots C_{x_m}^{\mathbf{p}^{(m)}}}(t)=\sum_{h \in \mathcal{A}_{Ax_1 \cdots x_m}}v_h(t)
  \label{equ:rem}
\end{equation}
holds for any $A$ and $x_i$ at time $t$. In the following, we prove Eq.~\ref{equ:rem} still holds at time $t+1$. Let's begin with proving that for each state $h \in \mathcal{A}_{Ax_1 \cdots x_m}$ $$\sum_{h' \in \mathcal{A}_{A'x'_1 \cdots x'_m}} m_{h,h'} = m''_{A C_{x_1}^{\mathbf{p}^{(1)}} \cdots C_{x_m}^{\mathbf{p}^{(m)}},A' C_{x'_1}^{\mathbf{p}^{(1)}} \cdots C_{x'_m}^{\mathbf{p}^{(m)}}}$$ for $A \in \{C,D\}$ and $x_i\in\{0,1,...,k_i\}$. For any states $h'  \in \mathcal{A}_{A'x'_1 \cdots x'_m}$, each cooperative player with $\mathbf{p}^{(i)}$ either cooperates or defects in the previous round. Suppose $j$ players with $\mathbf{p}^{(i)}$ in $h' \in \mathcal{A}_{A'x'_1 \cdots x'_m}$ cooperates in $h \in \mathcal{A}_{Ax_1 \cdots x_m}$. Then, the transition probability from $x_i$ cooperators with $\mathbf{p}$ to $x'_i$ cooperators is
\begin{equation*}
  \begin{aligned}
    &\tbinom{x_i}{j}  \left(p_{Cx+y-1+f(A)}^{(i)}\right)^{j}\left(1-p_{Cx+y-1+f(A)}^{(i)}\right)^{x_i-j}\\
    &\tbinom{k_i-x_i}{x'_i-j} \left(p_{Dx+y+f(A)}^{(i)}\right)^{x'_i-j}\left(1-p_{Dx+y+f(A)}^{(i)}\right)^{(k_i-x_i)-\left(x^{\prime}_i-j\right)}.
  \end{aligned}
\end{equation*}
There are at most $min\{x_i,x'_i\}$ cooperators in $h' \in \mathcal{A}_{Ax'_1 \cdots x'_m}$ who cooperate in $h \in \mathcal{A}_{Ax_1 \cdots x_m}$ and at least 0 cooperator taking action $C$ in $h \in \mathcal{A}_{Ax_1 \cdots x_m}$. We define $\tbinom{x_i}{x'_i}=0$ if $x'_i>x_i$, the sum of transition probabilities from each $h \in \mathcal{A}_{Ax_1 \cdots x_m}$ to all states in $h' \in \mathcal{A}_{A'x'_1 \cdots x'_m}$ is 
\begin{equation*}
  \begin{aligned}
    \left|1-f(A')-p_{Ax+y} ^{(i)}\right| \prod_{i=1}^m \sum_{j=0}^{x'_i} & \tbinom{x_i}{j}  \left(p_{Cx+y-1+f(A)}^{(i)}\right)^{j}\left(1-p_{Cx+y-1+f(A)}^{(i)}\right)^{x_i-j}\\
    &\tbinom{k_i-x_i}{x'_i-j} \left(p_{Dx+y+f(A)}^{(i)}\right)^{x'_i-j}\left(1-p_{Dx+y+f(A)}^{(i)}\right)^{(k_i-x_i)-\left(x^{\prime}_i-j\right)},
  \end{aligned}
\end{equation*}
which is equal to $m''_{A C_{x_1}^{\mathbf{p}^{(1)}} \cdots C_{x_m}^{\mathbf{p}^{(m)}},A' C_{x'_1}^{\mathbf{p}^{(1)}} \cdots C_{x'_m}^{\mathbf{p}^{(m)}}}$. 

Given the distribution of states at time $t$, a straightforward calculation is,
\begin{equation}
  \begin{aligned}
    \sum_{h\in\mathcal{A}_{Ax_1 \cdots x_m}} v_h(t+1) &= \sum_{h\in\mathcal{A}_{Ax_1 \cdots x_m}} \sum_{h'\in \mathcal{A}^n} v_{h'}(t)m_{h',h}\\
    &=\sum_{h\in\mathcal{A}_{Ax_1 \cdots x_m}}\sum_{A',x'_1,...,x'_m} \sum_{h'\in\mathcal{A}_{A'x'_1 \cdots x'_m}}v_{h'}(t)m_{h',h}\\
    &=\sum_{A',x'_1,...,x'_m}\sum_{h'\in\mathcal{A}_{A'x'_1 \cdots x'_m}}v_{h'}(t)\sum_{h\in\mathcal{A}_{Ax_1 \cdots x_m}}m_{h',h}\\
    &=\sum_{A',x'_1,...,x'_m}\sum_{h'\in\mathcal{A}_{A'x'_1 \cdots x'_m}}v_{h'}(t)m''_{A' C_{x'_1}^{\mathbf{p}^{(1)}} \cdots C_{x'_m}^{\mathbf{p}^{(m)}},A C_{x_1}^{\mathbf{p}^{(1)}} \cdots C_{x_m}^{\mathbf{p}^{(m)}}}\\
    &=\sum_{A',x'_1,...,x'_m}\nu_{A' C_{x'_1}^{\mathbf{p}^{(1)}} \cdots C_{x'_m}^{\mathbf{p}^{(m)}}}(t)m''_{A' C_{x'_1}^{\mathbf{p}^{(1)}} \cdots C_{x'_m}^{\mathbf{p}^{(m)}},A C_{x_1}^{\mathbf{p}^{(1)}} \cdots C_{x_m}^{\mathbf{p}^{(m)}}}\\
    &=\nu_{A C_{x_1}^{\mathbf{p}^{(1)}} \cdots C_{x_m}^{\mathbf{p}^{(m)}}}(t+1).
  \end{aligned}
\end{equation}
Thus, $\nu_{A C_{x_1}^{\mathbf{p}^{(1)}} \cdots C_{x_m}^{\mathbf{p}^{(m)}}}(t)=\sum_{h \in \mathcal{A}_{Ax_1 \cdots x_m}}v_h(t)$ holds for any time $t$.

For finitely repeated multiplayer games, the mean distribution $\nu_{A C_{x_1}^{\mathbf{p}^{(1)}} \cdots C_{x_m}^{\mathbf{p}^{(m)}}}$ is give by
\begin{equation}
  \begin{aligned}
    \nu_{A C_{x_1}^{\mathbf{p}^{(1)}} \cdots C_{x_m}^{\mathbf{p}^{(m)}}}
    &=(1-\delta) \sum_{t=0}^{\infty} \delta^t \nu_{A C_{x_1}^{\mathbf{p}^{(1)}} \cdots C_{x_m}^{\mathbf{p}^{(m)}}}(t)\\
    &= (1-\delta) \sum_{t=0}^{\infty} \sum_{h \in \mathcal{A}_{Ax_1 \cdots x_m}} \delta^t v_h(t)\\
    &= (1-\delta) \sum_{h \in \mathcal{A}_{Ax_1 \cdots x_m}} \sum_{t=0}^{\infty} \delta^t v_h(t)\\
    &= \sum_{h \in \mathcal{A}_{Ax_1 \cdots x_m}} v_h.
  \end{aligned} 
\end{equation}
For infinitely repeated games, $\mathbf{\nu}$ is the unique left eigenvector of $\mathbf{M}''$. For $\epsilon >0$ the matrix $\mathbf{M}''$ is stochastic and primitive. Hence, $\mathbf{\nu}_{A C_{x_1}^{\mathbf{p}^{(1)}} \cdots C_{x_m}^{\mathbf{p}^{(m)}}} = \lim_{t \rightarrow \infty} \mathbf{\nu}_{A C_{x_1}^{\mathbf{p}^{(1)}} \cdots C_{x_m}^{\mathbf{p}^{(m)}}}(t)$ such that
\begin{equation}
  \nu_{A C_{x_1}^{\mathbf{p}^{(1)}} \cdots C_{x_m}^{\mathbf{p}^{(m)}}}=\sum_{h \in \mathcal{A}_{Ax_1 \cdots x_m}} v_h
\end{equation}
holds. 

\end{proof}

\newpage
\begin{figure}[h]
  \centering
  \includegraphics[width=5in]{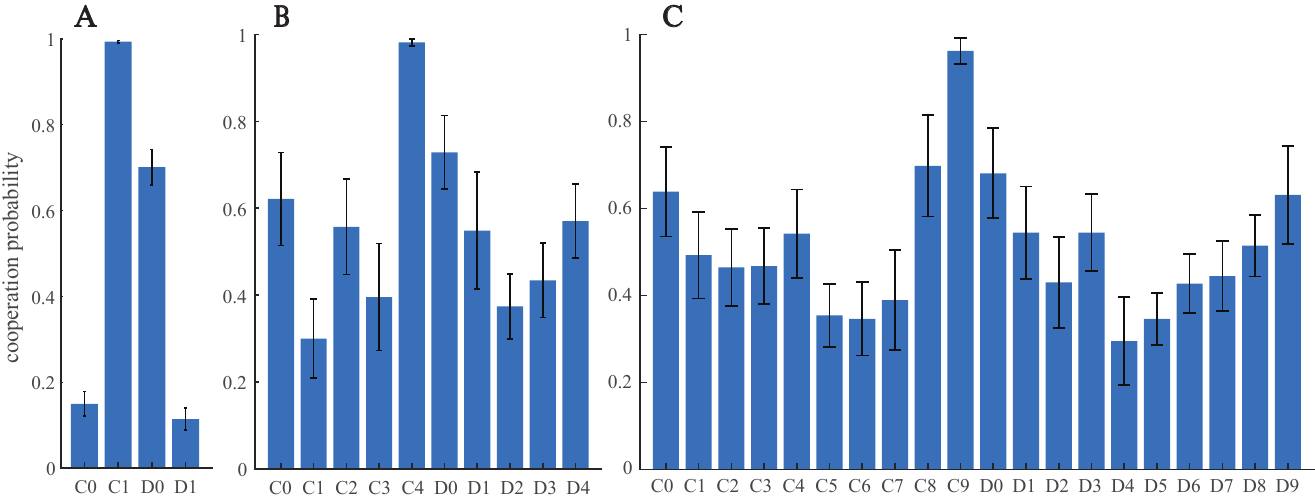}
  \caption{\textbf{The most abundant strategy in the infinitely repeated multiplayer social dilemma.} Apart from analyzing the average strategies of simulations in Fig. 2 of the main text, we also record the most abundant strategies. Subgraphs A, B, and C show the most abundant strategies in infinitely repeated social dilemmas with 2, 5, and 10 players, respectively. The error bars displays the standard error of 10 independent runs. Comparing to the most abundant strategies, the average strategies are more centralized, suggesting it's more accurate to study the evolution of cooperation by the average strategies. Similar to the average strategies, the most abundant strategies also have the property: cooperate if most players cooperate or defect, and defect if the medium number of players cooperate. Parameters are the same as Fig. 2 of the main text.}
  \label{fig:infinite_most}
\end{figure}

\begin{figure}
  \centering
  \includegraphics[width=5in]{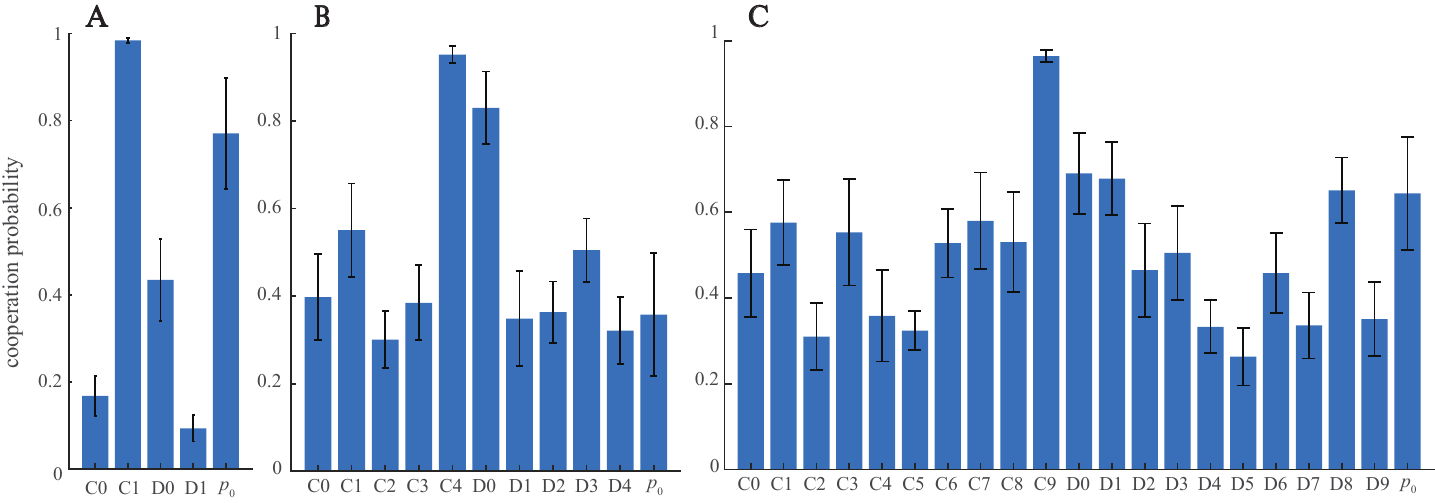}
  \caption{\textbf{The most abundant strategy in the finitely repeated multiplayer social dilemma.} We also record the most abundant strategies of the 10 independent simulations in Figure 2 of the main text. Subgraphs A, B, and C show the most abundant strategies in the finitely repeated social dilemmas with 2, 5, and 10 players, respectively. In all cases, players would cooperate if most players have either cooperated or defected. Players tend to defect if half of the players cooperated. The initial cooperative probability is near 0.5, implying initial cooperative probabilities have little influence on the survival of strategies if the time scale of evolution is large. The parameters are the same as Figure 2 of the main text.}
  \label{fig:finite_most}
\end{figure}

\begin{figure}
  \centering
  \includegraphics[width=5in]{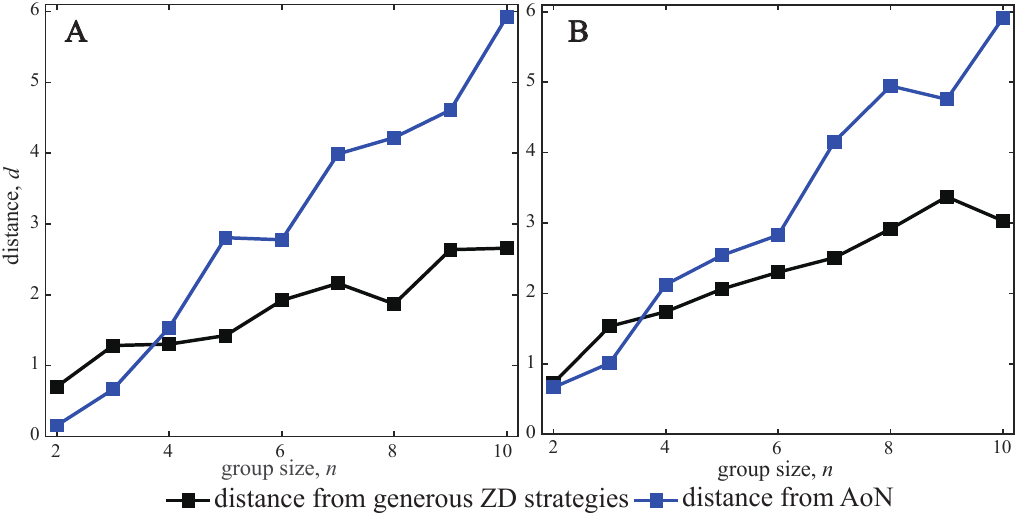}
  \caption{\textbf{Distances between the most abundant strategy and two important memory-1 strategies.} We further analyze the distance of the most abundant strategies and two of the most important memory-1 strategies: generous Zero-Determined (ZD) strategies and All-or-None (AoN) strategy. The distances are measured by Euclidean metric. (A) Results in the infinitely repeated games. (B) Results in the finitely repeated games. In both cases, the most abundant strategies are closer to generous ZD if the number of players is sufficiently large, which is the same as the average strategies. Thus, nature prefers generous ZD strategies to AoN. The parameters in A and B are the same as Figure 3 of the main text.}
  \label{fig:distance_most}
\end{figure}

\begin{figure}
  \centering
  \includegraphics[width=4.5in]{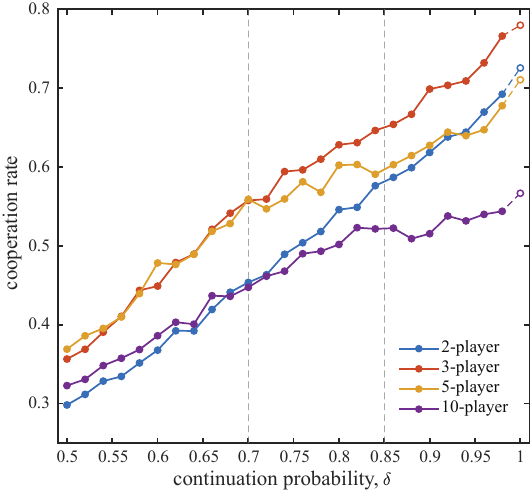}
  \caption{\textbf{Impact of continuation probability on cooperation rate in infinitely and finitely repeated multiplayer games.} For different continuation probability, we record the cooperation rates for different group size. We find more expected round leads to a higher cooperation rate, confirming that iterated interactions are favorable to the evolution of cooperation. Parameters: $c=1$, $r/n=0.85$, $N=100$, $\epsilon=0.01$ and $s=1$.}
  \label{fig:discount}
\end{figure}

\begin{figure}
  \centering
  \includegraphics[width=5in]{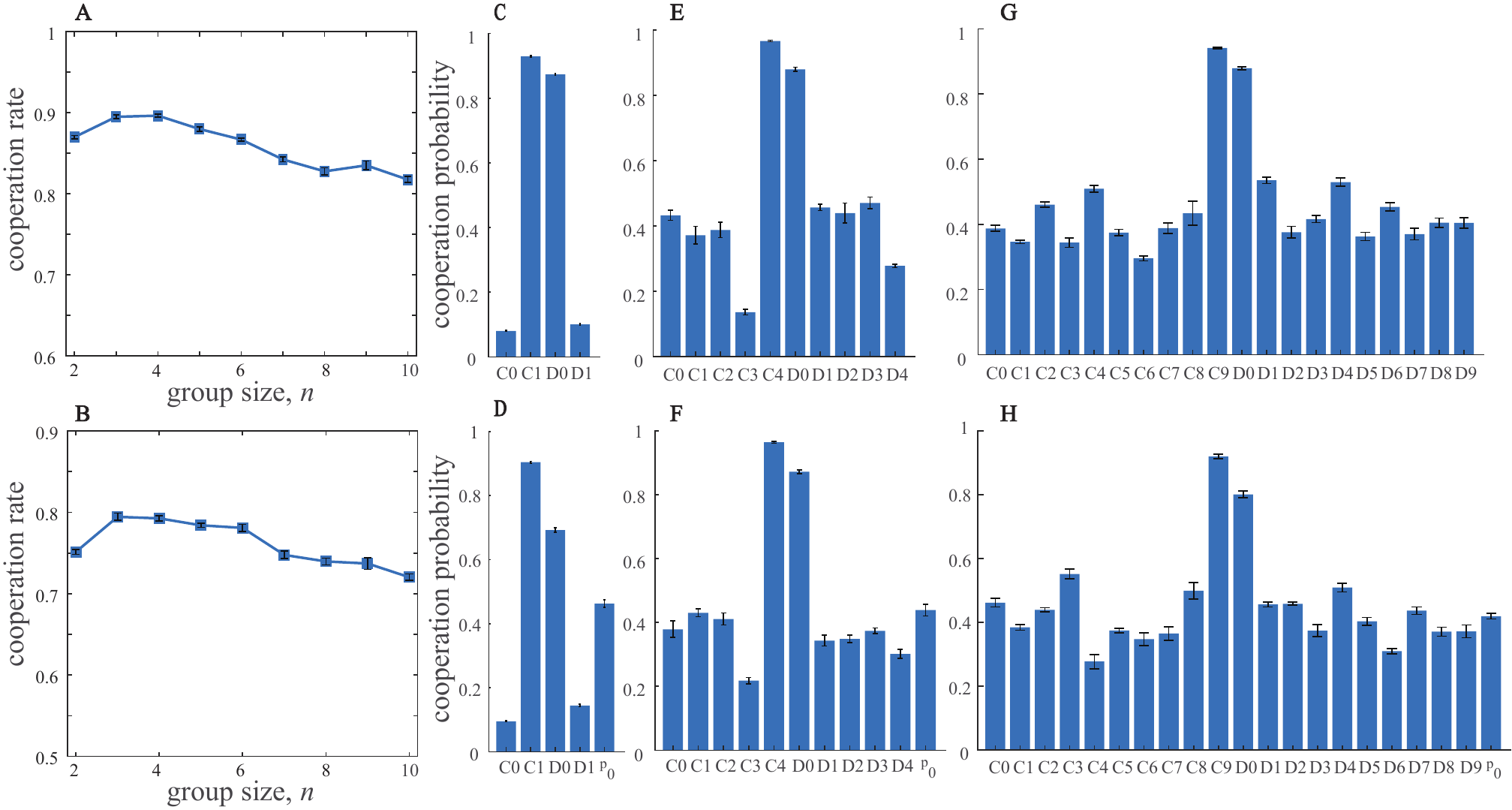}
  \caption{\textbf{Evolution of cooperation in repeated multiplayer games when strategies are drawn with a U-shaped density distribution.} Similar to Figure 2 of the main text, we consider two scenarios: infinitely repeated Public Goods Games and finitely repeated Public Goods Games with $\delta=0.85$. We also use the stochastic evolutionary dynamics of Imhof and Nowak to explore the evolution of cooperation, but assume that each bit of memory-1 strategies is sampled from a U-shaped density distribution $[\pi x(1-x)]^{-1/2}$ \cite{nowak1993strategy}. This distribution ensures that the randomly sampled strategy's elements are near 1 or 0. We introduce $10^6$ mutants to each simulation and execute 10 independent runs. (A, B) The impact of group size on cooperation rate. In both scenarios, as the group size increases the cooperation rate first increases and then decreases. (C-H) The average strategies during simulation for $n=2$ (C, D), $n=5$ (E, F) and $n=10$ (G, H). In both scenarios, players tend to cooperate if all players cooperate or all players defect in the previous round. Players shall defect otherwise. Although initial probabilities have non-negligible influence on payoffs in finitely repeated games, subgraphs D, F and H suggests players have no tendency to cooperating or defecting in the initial round, indicating the impact of initial probabilities can be ignored from an average point of view. Parameters  in this figure are the same as that in Figure 2 of the main text.}
  \label{fig:U-shape}
\end{figure}

\end{spacing}
\end{document}